\documentclass{SciPost}

\hypersetup{
    colorlinks,
    linkcolor={red!50!black},
    citecolor={blue!50!black},
    urlcolor={blue!80!black}
}

\usepackage[bitstream-charter]{mathdesign}
\DeclareSymbolFont{usualmathcal}{OMS}{cmsy}{m}{n}
\DeclareSymbolFontAlphabet{\mathcal}{usualmathcal}

\fancypagestyle{SPstyle}{
\fancyhf{}
\lhead{\colorbox{scipostblue}{\bf \color{white} ~SciPost Physics }}
\rhead{{\bf \color{scipostdeepblue} ~Submission }}

\fancyfoot[C]{\textbf{\thepage}}
}

\usepackage{booktabs}
\usepackage{hyperref}
\usepackage[percent]{overpic}
\usepackage{caption}
\usepackage{xspace}

\usepackage{array}
\usepackage{pdflscape}

\newcommand{\eg}{\mbox{e.g.}\xspace}

\newcommand{\ie}{\mbox{i.e.}\xspace}
\newcommand{\alps}{\ensuremath{\alpha_s}\xspace}

\newcommand{\cnp}{\ensuremath{C_\text{NP}}\xspace}
\newcommand{\chad}{\ensuremath{C_\text{Had}}\xspace}
\newcommand{\cmpi}{\ensuremath{C_\text{MPI}}\xspace}
\newcommand{\etaabs}{\ensuremath{\vert \eta\vert}\xspace}
\newcommand{\kt}{\ensuremath{k_\text{T}}\xspace}
\newcommand{\mjj}{\ensuremath{m_{\text{1,2}}}\xspace}
\newcommand{\mz}{\ensuremath{m_{\text{Z}}}\xspace}
\newcommand{\pt}{\ensuremath{p_{\text{T}}}\xspace}
\newcommand{\ptjet}{\ensuremath{p_{\text{T,jet}}}\xspace}
\newcommand{\ptone}{\ensuremath{p_{\text{T,jet1}}}\xspace}
\newcommand{\pttwo}{\ensuremath{p_{\text{T,jet2}}}\xspace}

\newcommand{\ptave}{\ensuremath{\langle{}p_\text{T}\rangle_\text{1,2}}\xspace}

\newcommand{\ptz}{\ensuremath{p_\text{T,Z}}\xspace}

\newcommand{\yjetone}{\ensuremath{y_{\text{jet1}}}\xspace}
\newcommand{\yone}{\ensuremath{y_{\text{1}}}\xspace}
\newcommand{\ytwo}{\ensuremath{y_{\text{2}}}\xspace}
\newcommand{\yz}{\ensuremath{y_{\text{Z}}}\xspace}
\newcommand{\yboost}{\ensuremath{y_\text{b}}\xspace}
\newcommand{\yabsmax}{\ensuremath{\left|y\right|_{\text{max}}}\xspace}
\newcommand{\ystar}{\ensuremath{y^*}\xspace}
\newcommand{\ybys}{\ensuremath{(\yboost,\ystar)}\xspace}
\newcommand{\fastjet}{\textsc{FastJet}\xspace}
\newcommand{\herwig}{\textsc{Herwig7}\xspace}
\newcommand{\law}{\textsc{law}\xspace}
\newcommand{\mcrun}{\textsc{MCRun}\xspace}
\newcommand{\rivet}{\textsc{Rivet}\xspace}
\newcommand{\sherpa}{\textsc{Sherpa}\xspace}

\newcommand{\GeV}{\ensuremath{\,\text{GeV}}\xspace}
\newcommand{\MeV}{\ensuremath{\,\text{MeV}}\xspace}
\newcommand{\hftwo}{\hspace*{\fill}}
\begin{document}

\pagestyle{SPstyle}

\begin{center}{\Large \textbf{\color{scipostdeepblue}{
        Nonperturbative effects in triple-differential dijet and Z+jet production at the LHC\\}}}
\end{center}

\begin{center}\textbf{
    Stefan~Gieseke\textsuperscript{3$\dagger$\mbox{\,\href{https://orcid.org/0000-0003-4058-0954}{\includegraphics[scale=0.075]{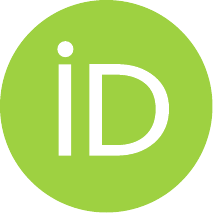}}}},
    Maximilian~Horzela\textsuperscript{1\mbox{\,\href{https://orcid.org/0000-0002-3190-7962}{\includegraphics[scale=0.075]{orcid.pdf}}}},
    Manjit~Kaur\textsuperscript{4\mbox{\,\href{https://orcid.org/0000-0002-3440-2767}{\includegraphics[scale=0.075]{orcid.pdf}}}},
    Dari~Leonardi\textsuperscript{3},
    Klaus~Rabbertz\textsuperscript{2$\star$\mbox{\,\href{https://orcid.org/0000-0001-7040-9846}{\includegraphics[scale=0.075]{orcid.pdf}}}},
    Aayushi~Singla\textsuperscript{4\mbox{\,\href{https://orcid.org/0000-0003-2550-139X}{\includegraphics[scale=0.075]{orcid.pdf}}}} and
    Cedric~Verstege\textsuperscript{2\mbox{\,\href{https://orcid.org/0000-0002-2816-7713}{\includegraphics[scale=0.075]{orcid.pdf}}}}}
\end{center}

\begin{center}
  {\bf 1} II.~Institute of Physics, Georg-August-Universität Göttingen, Friedrich-Hund-Platz~1, 37077 Göttingen, Germany\\
  {\bf 2} Institute of Experimental Particle Physics, Karlsruhe Institute of Technology, Wolfgang-Gaede-Str.~1, 76131 Karlsruhe, Germany\\
  {\bf 3} Institute of Theoretical Physics, Karlsruhe Institute of Technology, Wolfgang-Gaede-Str.~1, 76131 Karlsruhe, Germany\\
  {\bf 4} Department of Physics, Panjab University, Sector-14, 160014 Chandigarh, India
  \\[\baselineskip]
  $\star$ \href{mailto:klaus.rabbertz@kit.edu}{\small klaus.rabbertz@kit.edu}\,,\quad
  $\dagger$ \href{mailto:stefan.gieseke@kit.edu}{\small stefan.gieseke@kit.edu}
\end{center}

\section*{\color{scipostdeepblue}{Abstract}}
\textbf{\boldmath{%
    In comparisons of precision measurements at colliders to the most accurate predictions available in perturbative quantum chromodynamics (QCD), it
    is required to correct for nonperturbative effects. By means of Monte Carlo event generators this article investigates the impact of such
    nonperturbative effects on two processes relevant for precision determinations of the strong coupling constant and the proton structure:
    triple-differential dijet and Z+jet production. We observe significant differences between the two processes. Whether this non-universal behaviour
    is realized in nature remains an open question. We therefore propose a triple-differential measurement of the underlying event in Z+jet
    production.}}

\vspace{\baselineskip}

\noindent\textcolor{white!90!black}{%
\fbox{\parbox{0.975\linewidth}{%
\textcolor{white!40!black}{\begin{tabular}{lr}%
  \begin{minipage}{0.6\textwidth}%
    {\small Copyright attribution to authors. \newline
    This work is a submission to SciPost Physics. \newline
    License information to appear upon publication. \newline
    Publication information to appear upon publication.}
  \end{minipage} & \begin{minipage}{0.4\textwidth}
    {\small Received Date \newline Accepted Date \newline Published Date}%
  \end{minipage}
\end{tabular}}
}}
}


\vspace{10pt}
\noindent\rule{\textwidth}{1pt}
\tableofcontents
\noindent\rule{\textwidth}{1pt}
\vspace{10pt}
\clearpage


\section{Introduction}
\label{sec:intro}

To match the unprecedented precision of measurements at the LHC, accurate predictions of at least next-to-next-to-leading order (NNLO) in perturbative
QCD (pQCD) are required. An indispensable ingredient for such calculations are parton distribution functions (PDFs) that describe the manifestly
nonperturbative (NP) internal structure of the colliding hadrons and in particular of the proton. Despite significant progress in lattice gauge theory
it is not yet possible to derive the proton structure from first principles such that the proton PDFs must be determined in dedicated fits of
PDF-dependent observables to precise measurements. Moreover, further NP effects from hadronisation and the Underlying Event (UE) must be estimated by
means of Monte Carlo (MC) event generators, since typically the highest-order predictions in pQCD only provide events at parton level.

\begin{figure}[htbp]
  \hftwo\includegraphics[width=0.30\linewidth]{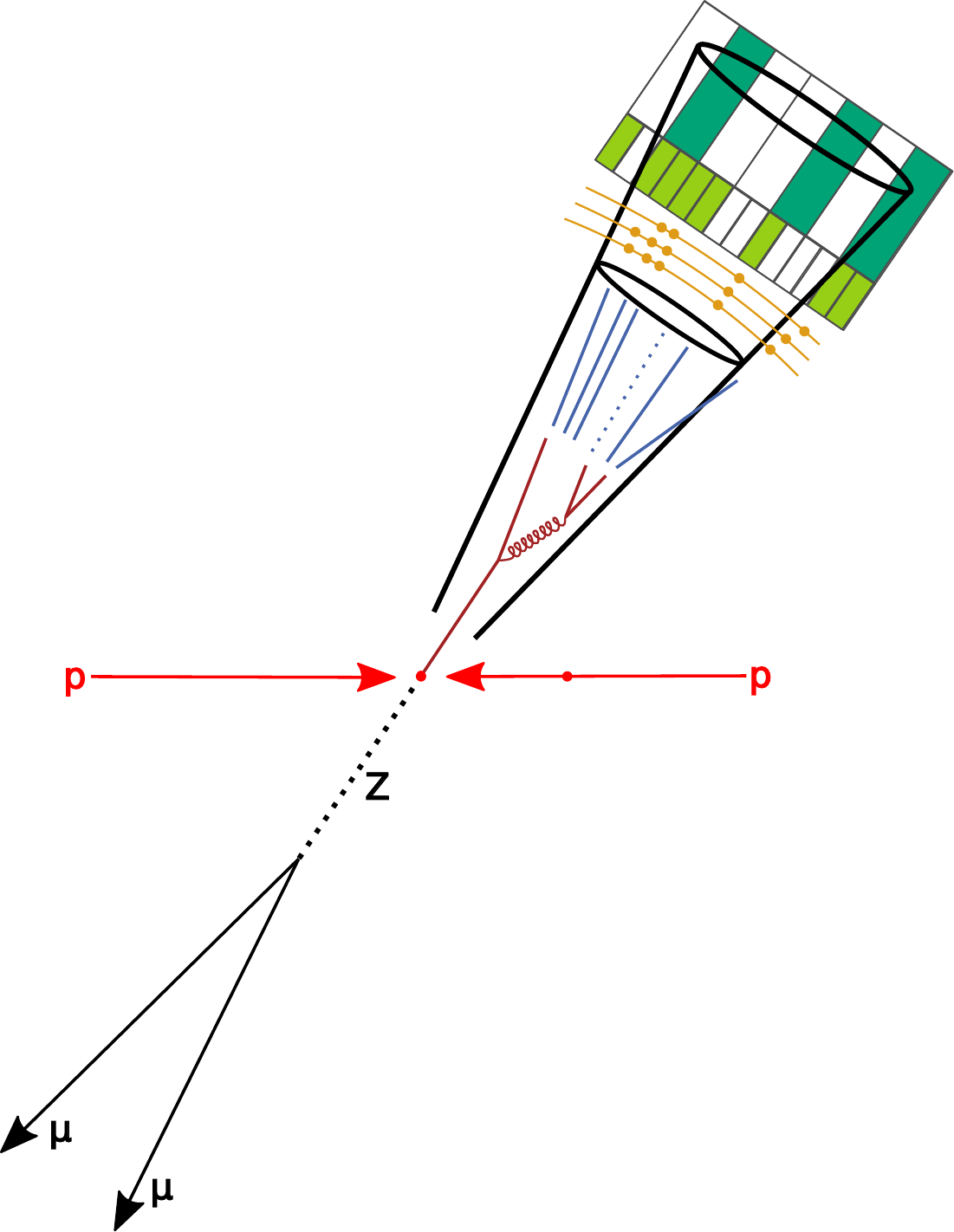}\hftwo%
  \includegraphics[width=0.30\linewidth]{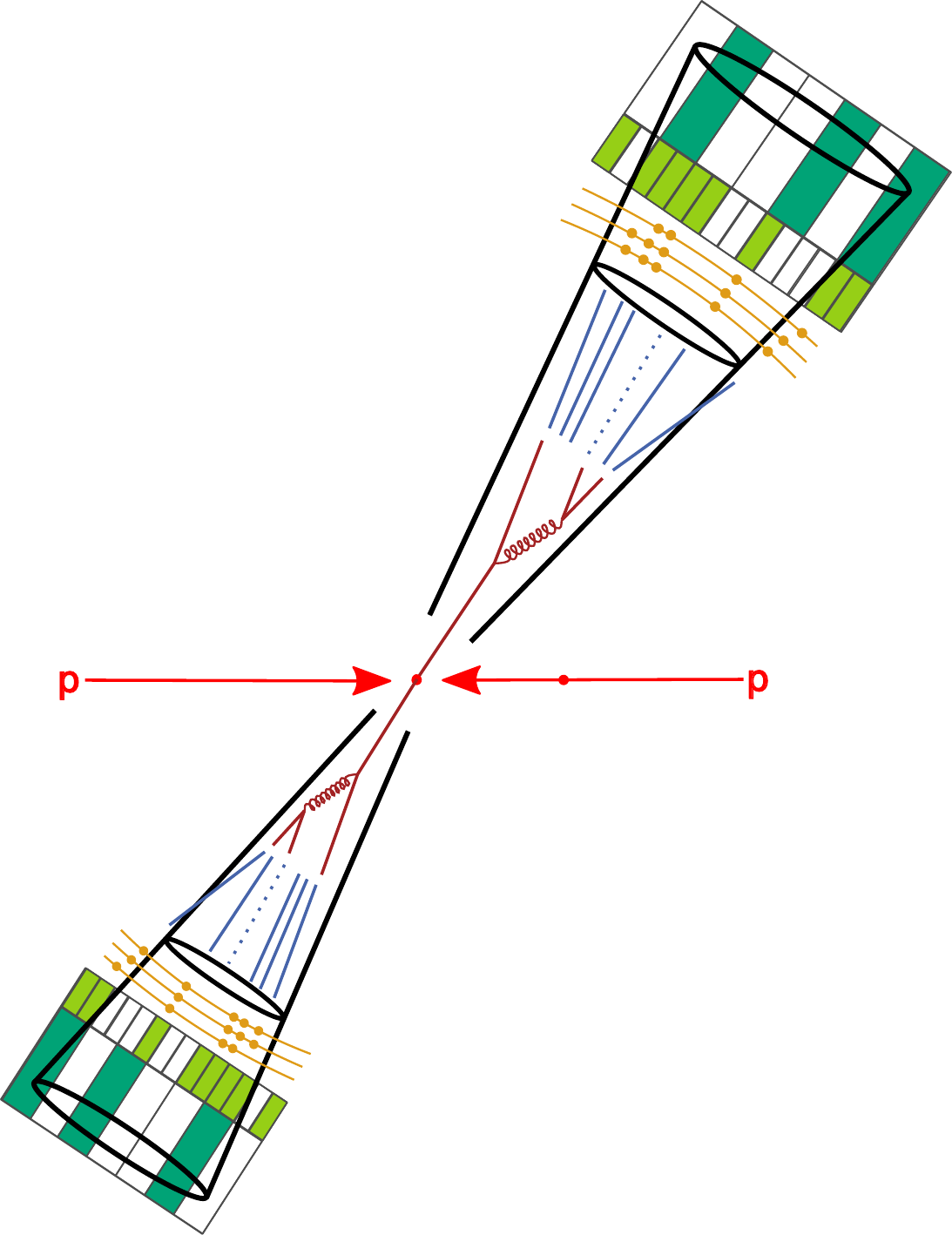}\hftwo
  \caption{Illustrations of Z+jet (left)~\cite{Horzela2023} and dijet production (right). In the Z+jet case, an oppositely-charged muon pair and at
    least one jet are required. For the dijet case, a second jet replaces the muon pair.}
  \label{fig:topology}
\end{figure}

In this article, NP effects for two processes in proton--proton collisions at the LHC are studied: the inclusive production of dijet events,
$\,p + p \rightarrow \mathrm{jet} + \mathrm{jet} + X$, and, generically labelled as Z+jet, the process
$\,p + p \rightarrow Z/\gamma^* + \mathrm{jet} + X$, where only the decay channel $Z/\gamma^*\rightarrow\mu^+\mu^-$ is considered. %
For both processes, illustrated in Fig.~\ref{fig:topology}, predictions are available at NNLO accuracy in
pQCD~\cite{Boughezal:2015ded,Ridder:2016nkl,Currie:2017eqf,Czakon:2019tmo,Chen:2022tpk}.  While baseline fits of proton PDFs rely on deep-inelastic
scattering (DIS)~\cite{Abramowicz:2015mha}, which to first degree depends on the valence quarks and only indirectly on the strong coupling \alps,
dijet and Z+jet production both are directly sensitive to \alps and the gluon content of the
proton~\cite{Ablat:2024uvg,Hou:2019efy,NNPDF:2021njg,Cridge:2023ozx}. Measurements of these processes at the LHC therefore hold significant potential
towards improving the accuracy of \alps and the gluon PDF{}.

Following the publications by the CMS Collaboration on dijet production at the LHC~\cite{CMS:2017jfq,CMS:2023fix} we choose a setup where the cross
section is investigated triple-differentially as a function of the two angular variables \yboost and \ystar plus one quantity with dimension of energy
that correlates with the energy scale of the collision. The two angular variables are defined as
\begin{equation}
  \yboost = \frac{1}{2} \left| \yone + \ytwo \right|
  \label{eq:yboost}
\end{equation}
and
\begin{equation}
  \ystar = \frac{1}{2} \left| \yone - \ytwo \right|
  \label{eq:ystar}
\end{equation}
where \yone and \ytwo are the rapidities of the two leading-\pt objects emerging from a collision, \ie either the leading two jets in dijet production
or the reconstructed Z~boson and the leading jet in Z+jet events. The quantity \yboost corresponds to the longitudinal boost of the centre-of-mass
with respect to the laboratory system of a $2 \rightarrow 2$ scatter, while \ystar is the rapidity of the two back-to-back final-state objects in the
centre-of-mass frame.  This choice is particularly useful, because the variables \ystar and \yboost to first order separate the dependence on the
partonic scattering angle in the centre-of-mass system from the dependence on the imbalance in the initial-state parton momenta, and hence the PDFs.
Moreover, one can derive the following relations to the proton momentum fractions $x_{1|2}$ carried by the incoming partons of a $2 \rightarrow 2$
scatter:
\begin{eqnarray}
  x_{1|2} &=& \frac{2\ptave~\text{cosh}(\ystar)}{\sqrt{s}}\exp(\pm \yboost)\nonumber\\
          &=& \frac{\mjj}{\sqrt{s}}\exp(\pm \yboost)\,.\label{eq:x1x2}
\end{eqnarray}
Here, \ptave and \mjj are the average transverse momentum, $\ptave = (\ptone+\pttwo)/2$, and the invariant mass,
$\mjj=\sqrt{({\bf p}_1+{\bf p}_2)^2}$, of the leading two jets, where \ptone, \pttwo are their transverse momenta and ${\bf p}_1$, ${\bf p}_2$ their
four-momenta, respectively.  Since at leading order (LO) for dijet events $\ptone = \pttwo = \ptave = \ptjet$ and for Z+jet events $\ptz = \ptjet$ we
focus in the following on \ptave and \ptz as third variable, which we generically label as $X$.  Furthermore, jets are always clustered with the
collinear- and infrared-safe anti-\kt jet algorithm~\cite{Cacciari:2008gp} with a size parameter of $R = 0.4$. %
Results with respect to the dijet mass, \ie $X=\mjj$, or for a large jet size parameter of $R = 0.8$ can be found as additional material in
Appendix~\ref{sec:addmat}.

The triple-differential cross section in these three observables can be written in terms of weighted MC events as
\begin{equation}
  \frac{d^{3}\sigma}{dX{}d\yboost{}d\ystar{}}=\frac{N_\mathrm{eff}\left(\Delta{}X,\Delta\yboost,\Delta\ystar\right)}{N_\mathrm{eff,tot}}\sigma_\mathrm{incl}\,,
  \label{eq:xsec}
\end{equation}
where $N_\mathrm{eff}\left(\Delta{}X,\Delta\yboost,\Delta\ystar\right)$ denotes the effective number of selected events within a given phase space in
$X$, \yboost, and \ystar; $N_\mathrm{eff,tot}$ is the effective number of generated events in the full phase space and $\sigma_\mathrm{incl}$ is the
inclusive cross section for the analysed process. The effective number of events can be expressed in terms of event weights $w_i$ as
\begin{equation}
  N_\mathrm{eff,\mathcal{S}} = \sum_{i \in \mathcal{S}} w_i
\end{equation}
of the events $i$ within a phase space region $\mathcal{S}$.

This article is organised as follows: %
Section~\ref{sec:phasespace} defines the phase space for the analyses. Nonperturbative corrections for the two investigated processes are derived and
discussed in Sect.~\ref{sec:npcorr}. The separation of NP effects into hadronisation and the Underlying Event is addressed in
Sect.~\ref{sec:hadmpicorr}. A dedicated section~\ref{sec:ue} examines the behaviour of typical observables for UE studies in the context of the
previous section. A summary of the main findings is presented in the concluding Sect.~\ref{sec:summary}.  Two appendices provide details on the
settings of the MC event generation, App.~\ref{sec:genconfig}, and additional results for $X=\mjj$ as observable or for $R=0.8$ as jet
size parameter, App.~\ref{sec:addmat}.

\section{Analysis phase space}
\label{sec:phasespace}

\begin{figure}[tbp]
  \centering%
  \includegraphics[width=0.5\linewidth]{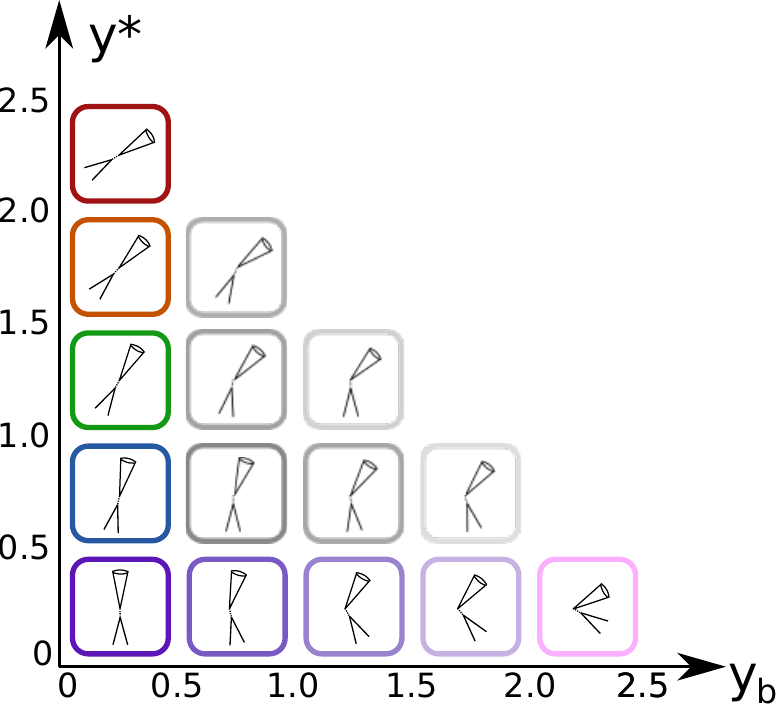}
  \caption{Visualisation of the analysed Z+jet phase space in \yboost and \ystar as defined in Eqs.~\ref{eq:yboost}-\ref{eq:ystar}. The bins along the
    \yboost and \ystar axes discussed in the main part of this work are indicated by coloured boxes.}
  \label{fig:ybyszj}
\end{figure}

The derived variables \yboost and \ystar of Eqs.~\ref{eq:yboost} and~\ref{eq:ystar} depend on the rapidities \yone, \ytwo of the leading objects
measured in a collision event. The latter are constrained by acceptance limitations in the experiments, \eg ATLAS or
CMS~\cite{Aad:2008zzm,Chatrchyan:2008aa}, which can be expressed in terms of the coverage in pseudorapidity $\eta$ of the relevant subdetectors. For
an optimal reconstruction, jets should be measured within coverage of the inner tracking devices extending typically up to $\etaabs\approx 2.5$. Muons
can be measured by the ATLAS (CMS) detectors up to $\etaabs = 2.7$ (2.4), leading to a similar constraint on the rapidity \yz of the Z~boson that is
reconstructed from an opposite-sign muon pair.  The transition from coordinate axes (\yone, \ytwo) to \ybys corresponds to a rotation by $\pi/4$ in
rapidity phase space. Applying cuts on (\yone, \ytwo) leads, after mapping to the first quadrant only, to a triangular-shaped phase space in \ybys, as
illustrated in Fig.~\ref{fig:ybyszj} for the Z+jet process. Here, we choose a binning scheme for \yboost and \ystar with a bin width of 0.5 starting
from zero up to 2.5. The maximum absolute rapidity allowed for either \yone or \ytwo is given by $\yboost + \ystar < \yabsmax = 3.0$, which is
slightly larger than the tracker coverage. This results in a total of 15 \ybys bins.

In the region where $\yboost < 0.5$, there are five bins ranging from $\ystar < 0.5$ to $2.0 \leq \ystar < 2.5$. Each following \yboost region has one
\ystar bin less than the previous one, finishing with a single bin $\ystar < 0.5$ for $2.0 \leq \yboost < 2.5$.  In the main part of this work, we
focus on the nine bins corresponding either to a series of bins with increasing \yboost for $\ystar<0.5$, or with increasing \ystar for $\yboost<0.5$
to better showcase our observations. These bins are indicated by coloured boxes in Fig.~\ref{fig:ybyszj}, while results for the remaining bins, drawn
with grey boxes, can be found in Figs.~\ref{fig:dijets_herwig_ak4_all}--\ref{fig:zjets_ak8_zpt_all} of Appendix~\ref{sec:addmat}.

For the final selection criteria and binning of dijet events we follow the publication by CMS~\cite{CMS:2023fix} and impose $\yabsmax = 3.0$ for the
two leading jets. In contrast to CMS, however, we reduce the \ptjet thresholds from $\ptone > 100\GeV$ and $\pttwo > 50\GeV$ to 30\GeV and 25\GeV
respectively. This allows for an increased overlap in phase space when comparing the dijet and Z+jet processes.

Since the production cross section drops drastically with increasing \ptave, the bin widths grow with \ptave to accommodate for the lower expected
number of events. Equally, towards outer rapidity the cross section becomes smaller. Therefore, bins in \ptave are merged in nine edge bins~(E) at
outer rapidities as compared to six central ones~(C). The final binning scheme chosen for dijet events is presented in Table~\ref{tab:dijet_binning}.

\begin{table}[tbp]
  \centering%
  \caption[The binning scheme in \ptave for the dijet analysis.]{The binning scheme in \ptave for the dijet analysis. Due to limited statistical
    precision, some bins are merged in the edge bins~(E) as compared to the central bins~(C) in the plane of \ybys.}
  \begin{tabular}{cl}
    \toprule
    Binning scheme & binning labels in \ybys plane\\
    \midrule
                   & \mbox{E}\\
                   & \mbox{E\quad{}E}\\
    (cf.~Fig.~\ref{fig:ybyszj}) & \mbox{C\quad{}E\quad{}E}\\
                   & \mbox{C\quad{}C\quad{}E\quad{}E}\\
                   & \mbox{C\quad{}C\quad{}C\quad{}E\quad{}E}\\
    \midrule
                   & bin edges in $\ptave\,/\GeV$\\
    \midrule
    Central~(C) & 47, 57, 67, 77, 87, 97, 107, 127,\\
                   & 147, 175, 207, 243, 284, 329, 380,\\
                   & 437, 499, 569, 646, 732, 827, 931,\\
                   & 1046, 1171, 1307, 1458, 1621, 1806,\\
                   & 2003, 2217, 2453, 2702 \\
    Edge~(E) & 47, 57, 67, 77, 87, 97, 107, 127,\\
                   & 147, 207, 284, 380, 499, 646, 827,\\
                   & 1046, 1307 \\
    \bottomrule
  \end{tabular}
  \label{tab:dijet_binning}
\end{table}

For the Z+jet process, the Z~boson is reconstructed from a pair of oppositely charged muons, where the invariant mass of the dimuon system has to be
within a range of $\pm 20\GeV$ around the Z-boson mass $\mz = 91.1876\GeV$~\cite{ParticleDataGroup:2022pth}. Muons are considered for an absolute
pseudorapidity smaller than $2.4$ and a transverse momentum of at least 25\GeV. If multiple Z-boson candidates can be reconstructed, the one closest
to \mz is chosen. The transverse momentum of the reconstructed Z~boson has to be at least $\ptz > 25\GeV$. Jets are required to have a minimum \pt of
20\GeV and an absolute rapidity of less than $2.4$.  Any jet that overlaps with a prompt muon within $\Delta R(\text{jet},\mu)<0.3$ is omitted.  The
distance $\Delta R$ to the muon is defined as
\begin{equation}
  \Delta R(\mu,\text{jet}) = \sqrt{\left(\phi_\text{jet}-\phi_\mu\right)^2 + \left(\eta_\text{jet}-\eta_\mu\right)^2}\,,\nonumber
\end{equation}
where $\phi$ and $\eta$ are the azimuthal angles and pseudorapidities of the corresponding jet and muon. This ensures that muons, which are also
clustered to form jets, are removed from further consideration as ``jet''.

Similar to the dijet case, a binning scheme with bin widths increasing with \ptz and fewer bins in the outer rapidity regions is chosen.  Moreover,
the suppression of the LO Z+jet process through PDF effects reduces the expected number of events at high \ystar, where parton momenta of
$x_1\approx x_2$ are required. As a result, an extra-wide binning~(X) is employed for the \ybys bin with ($\yboost<0.5$, $2.0\leq\ystar<2.5$). The
final binning scheme chosen for Z+jet events is presented in Table~\ref{tab:zj_binning}.

\begin{table}[tbp]
  \centering%
  \caption[The binning scheme in \ptz for the Z+Jet analysis.]{The binning scheme in \ptz for the Z+Jet analysis. Due to limited statistical
    precision, some bins are merged in the edge bins~(E) as compared to the central bins~(C) in the plane of \ybys. An extra wide binning~(X) is used
    for the highest \ystar-bin.}
  \begin{tabular}{cl}
    \toprule
    Binning scheme & binning labels in \ybys plane\\
    \midrule
                   & \mbox{X}\\
                   & \mbox{E\quad{}E}\\
    (cf.~Fig.~\ref{fig:ybyszj}) & \mbox{C\quad{}C\quad{}E}\\
                   & \mbox{C\quad{}C\quad{}C\quad{}E}\\
                   & \mbox{C\quad{}C\quad{}C\quad{}C\quad{}E}\\
    \midrule
                   & bin edges in $\ptz\,/\GeV$\\
    \midrule
    Central~(C) & 25, 30, 35, 40, 45, 50, 60, 70, 80, 90,\\
                   & 100, 110, 130, 150, 170, 190, 220,\\
                   & 250, 400, 1000\\
    Edge~(E) & 25, 30, 35, 40, 45, 50, 60, 70, 80, 90,\\
                   & 100, 110, 130, 150, 170, 190, 250,\\
                   & 1000\\
    Extra~(X) & 25, 30, 40, 50, 70, 90, 110, 150, 250 \\
    \bottomrule
  \end{tabular}
  \label{tab:zj_binning}
\end{table}

\section{Nonperturbative effects}
\label{sec:npcorr}

We derive correction factors to fixed-order perturbative calculations for the presented dijet and Z+jet analyses to estimate the size and uncertainty
of the effect introduced by the NP models in MC event generators. The full event generation until particle level usually passes through the following
steps: the calculation of the perturbative scattering amplitudes and subsequent phase space integration (matrix element, ME), the collinear
resummation of leading logarithms with parton showers (PS), the fragmentation of coloured final-state partons into hadrons and subsequent decays of
unstable hadrons (Had), and the addition of UE activity via multiple-parton interactions (MPI).

Similar as in previous analyses at the Tevatron or LHC~\cite{CDF:2007ykk,D0:2011jpq,CMS:2014qtp,CMS:2023fix,ATLAS:2025ifq}, the two steps labelled as
Had and MPI are considered as NP effects. Hence, we define the NP correction to fixed-order predictions at parton level as the ratio between the
nominal predictions with and without hadronisation effects and MPI. The correction at a given perturbative order (PO) of the ME can then be written
as:
\begin{equation}
  C_{\text{NP}}^{\text{PO}} = \frac{\sigma_{\text{ME+PS+Had+MPI}}^{\text{PO}}}{\sigma_{\text{ME+PS}}^{\text{PO}}}\,,
  \label{eq:nppo}
\end{equation}
where the subscripts ME+PS+Had+MPI and ME+PS indicate, which steps of the MC event generation have been executed.  In other words, the numerator
represents the nominal cross section at the particle level given by any MC event generator, while the denominator is the cross section derived from
the same MC generator with hadronisation effects and MPI turned off. Because the coloured partons after parton showers can interact with coloured
partons stemming from the UE modelling via MPI (colour reconnection), the event generation process must be run twice, once for the numerator and a
second time for the denominator leading to statistically independent event samples.  As a consequence it is not possible to apply a more elaborate
approach using migration matrices for the two levels, although it would certainly be interesting to follow the migration of jets in phase space.

We emphasize that this approach assumes independence of parton shower and hadronization, which certainly is a simplification.  There are implicit
correlations between parton shower and hadronization, as the parton shower cutoff is an upper scale for the hadronization.  When the parton shower is
terminated at high scales, the hadronization can compensate for softer radiation that has not been generated by the former, leading to approximately
the same total correction.  The separate investigation of the two effects, however, may include a residual effect of this correlation.  Nevertheless,
we follow the conventional procedure to quantify non-perturbative effects in experimental analyses to date.

For an automated and scalable generation of the MC events and subsequent analysis, the \mcrun software framework~\cite{horzela_2024_14530596} based on
the workflow management tool \law~\cite{marcel_rieger_2024_13952360} is used. The \law package implements and calls the tasks related to the necessary
code executions, automatically resolves their dependencies, and manages and distributes the computational payloads on available computing
resources. The configurations used for the results in this work are made available within the \mcrun software allowing for an efficient reproduction
of the same workflow.  A dedicated \rivet~\cite{Bierlich:2019rhm} routine for each process, respectively, analyses event records produced by the two
MC event generators \herwig version~7.2.3~\cite{Bahr:2008pv,Bellm:2015jjp} and \sherpa version~2.2.15~\cite{Sherpa:2019gpd}.  An overview of the
generator settings for all samples generated with \herwig is given in Table~\ref{tab:herwiggensettings} of Appendix~\ref{sec:genconfigherwig} and in
Table~\ref{tab:sherpagensettings} of Appendix~\ref{sec:genconfigsherpa} for the \sherpa samples. %
Jets are clustered utilising the \fastjet~\cite{Cacciari:2008gp,Cacciari:2011ma} library. Selection cuts based on reconstructed objects, as described
in the previous Sect.~\ref{sec:phasespace}, are applied, and fiducial cross sections for each differential bin are calculated.  These cross sections
are computed for event generations up to particle (full) and parton level only (partial).  Finally, NP correction factors are calculated following
Eq.~\ref{eq:nppo} from the obtained cross sections for each bin.

The obtained correction factors \cnp are subject to statistical fluctuations due to the limited number of generated events in both the full and
partial generation sequences. To smooth these fluctuations, the NP correction is parameterised using an interpolating function
\begin{equation}
  f(x) = a \cdot \ln(x/\text{GeV})^b + c\,,
  \label{eq:npfitfunc}
\end{equation}
in each \ybys bin.  This parameterisation is chosen, since the NP effects are expected to contribute in phase-space regions where the typical energy
of the collision approaches hadronic energy scales while fading out towards higher energies. Since the third observable is correlated with the energy
scale of the collision process, we expect to see larger NP effects towards lower values of $X$, while the NP correction factors should approach unity
towards high $X$.  Indeed, the fitted values for the parameter $c$ are compatible with unity and the obtained fits match the MC simulation well
supporting these expectations independent of the analysed topology, perturbative order, or \ybys bin.

\subsection{Dijet production}
\label{subsec:npdijet}

\begin{figure*}[tbp]
  \centering%
  \includegraphics[width=0.49\linewidth]{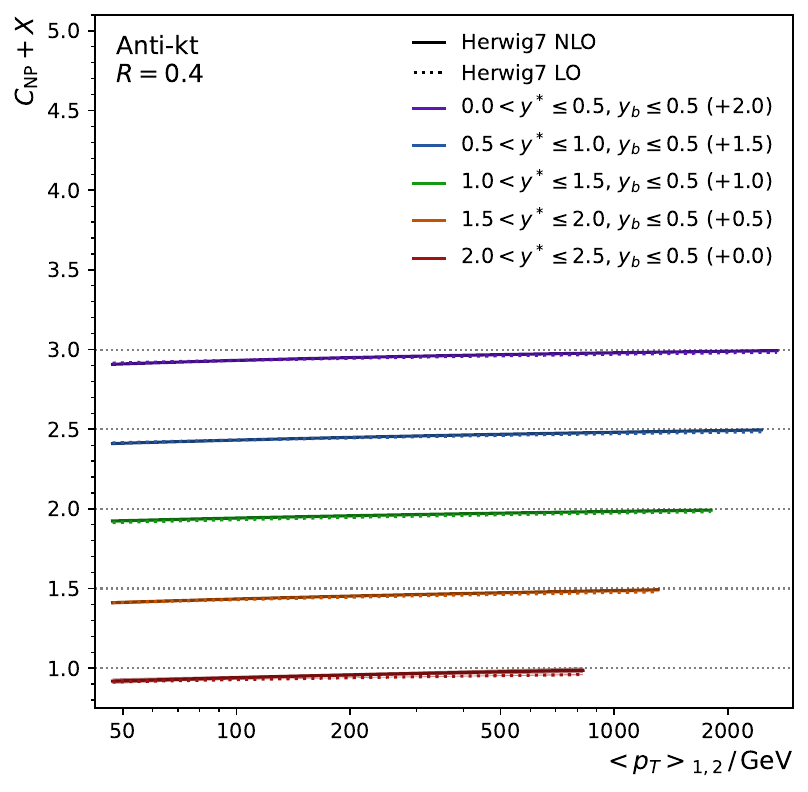}\hfill%
  \includegraphics[width=0.49\linewidth]{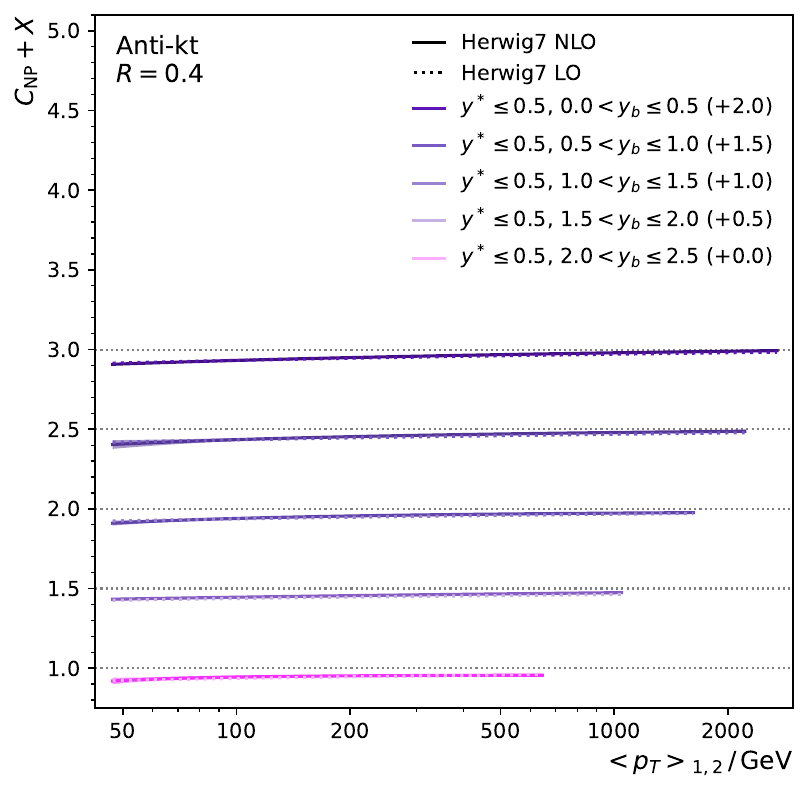}
  \caption{\cnp vs.\ dijet \ptave using LO (dotted) and NLO MEs (solid lines) in \herwig predictions for anti-\kt jets with $R=0.4$. Bands indicate
    the statistical uncertainty. The fitted curves are shown for a series of phase space intervals increasing in \ystar for $\yboost < 0.5$ (left) and
    increasing in \yboost for $\ystar < 0.5$ (right). The factors are shifted with respect to unity by constant offsets.}
  \label{fig:dijetAK4-np-herwig-lo-nlo}
\end{figure*}

The triple-differential dijet cross section is measured as a function of the three observables \yboost and \ystar, as defined in
Eqs.~\ref{eq:yboost}-\ref{eq:ystar}, and either the average transverse momentum \ptave or the mass \mjj of the dijet system.  The resulting NP
correction factors derived with \herwig at LO and next-to-leading order (NLO)~\cite{Bahr:2008pv,Bellm:2015jjp,Bewick:2023tfi,Platzer:2011bc} accuracy
in pQCD of the ME are shown in Fig.~\ref{fig:dijetAK4-np-herwig-lo-nlo} for anti-\kt jets with $R=0.4$ in the selected \ybys bins.  Further bins can
be found in Appendix~\ref{sec:addmat} in Fig.~\ref{fig:dijets_herwig_ak4_all} together with correction factors for dijet production vs.\ the dijet
invariant mass \mjj of the leading and subleading jet.  NP correction factors for anti-\kt jets with radius parameter $R=0.8$ can be found in
Fig.~\ref{fig:dijets_herwig_ak8_all} of the same appendix.

The observed correction factors for dijet production are close to unity for large \ptave (\mjj) and decrease towards small \ptave (\mjj) as
expected. They exhibit no significant dependence on the perturbative order or \ybys bin.

\subsection{Z+jet production}
\label{subsec:npzjet}

This process can be measured triple-differentially in a similar fashion as the dijet production. The observable that quantifies the scale of the hard
interaction is chosen to be the transverse momentum of the muon pair, labelled as \ptz in the following:
\begin{equation}
  \ptz = \left| \vec{p}_{\text{T},\mu^-} + \vec{p}_{\text{T},\mu^+} \right|\,,
\end{equation}
where $\vec{p}_{\text{T},\mu^-}$ and $\vec{p}_{\text{T},\mu^+}$ are the transverse momenta of the muon pair forming the $Z$-boson candidate.  The
rapidity observables \yboost and \ystar are defined analogously to Eqs.~\ref{eq:yboost}-\ref{eq:ystar} with \yone being the rapidity of the Z boson,
\yz, and \ytwo corresponding to the rapidity of the leading jet in the event \yjetone.

\begin{figure*}[tbp]
  \centering%
  \includegraphics[width=0.49\linewidth]{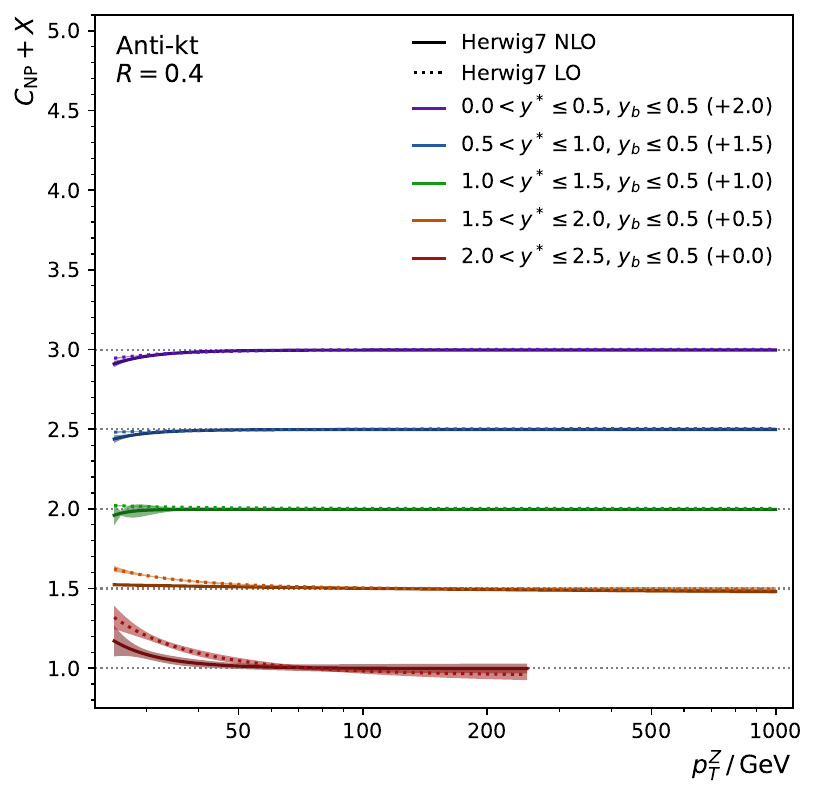}\hfill%
  \includegraphics[width=0.49\linewidth]{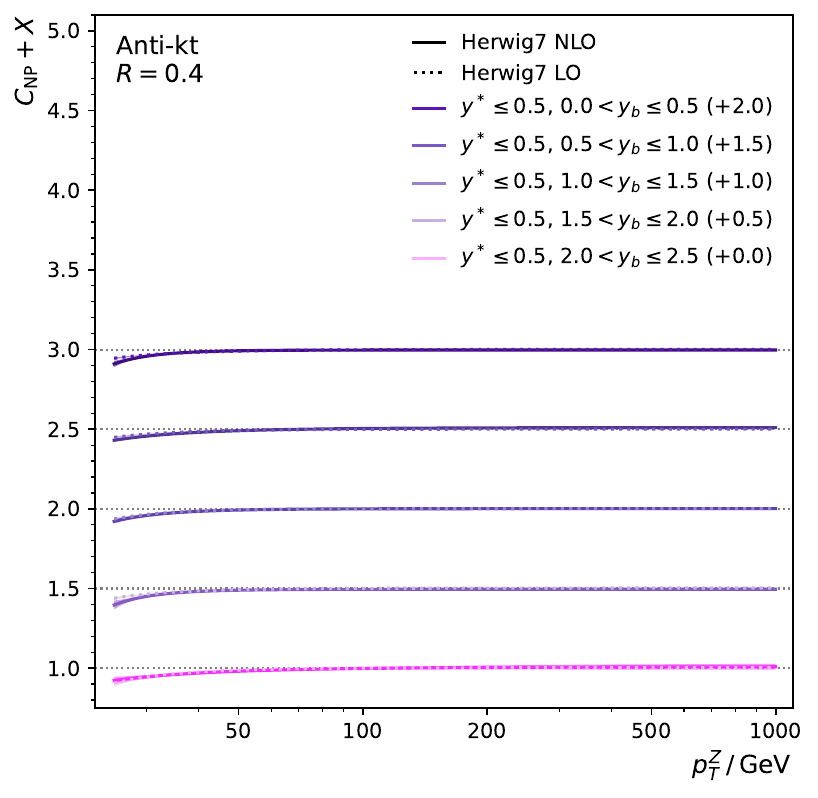}
  \caption{Same as Fig.~\ref{fig:dijetAK4-np-herwig-lo-nlo} but vs.\ \ptz in Z+jet production.}
  \label{fig:zjetAK4-np-herwig-lo-nlo}
\end{figure*}

NP correction factors derived with \herwig at LO and NLO accuracy in pQCD%
\footnote{Here, leading and next-to-leading order explicitly require the production of at least one jet in addition to the Z boson: Z+jet+X.} %
are presented in Fig.~\ref{fig:zjetAK4-np-herwig-lo-nlo} for anti-\kt jets with $R = 0.4$ for the selected \ybys bins.  Further \ybys bins as well as
all bins for anti-\kt jets with $R = 0.8$ can be found in Figs.~\ref{fig:zjets_ak4_zpt_all}--\ref{fig:zjets_ak8_zpt_all} in
Appendix~\ref{sec:addmat}.  In contrast to the behaviour observed for dijet production, we find a trend towards larger correction factors with rising
\ystar but not with rising \yboost. This trend in \ystar seems to be somewhat reduced using NLO MEs, where all NP correction factors exhibit an
overall decrease in magnitude as compared to LO MEs\@.

\begin{figure*}
  \centering%
  \includegraphics[width=0.49\linewidth]{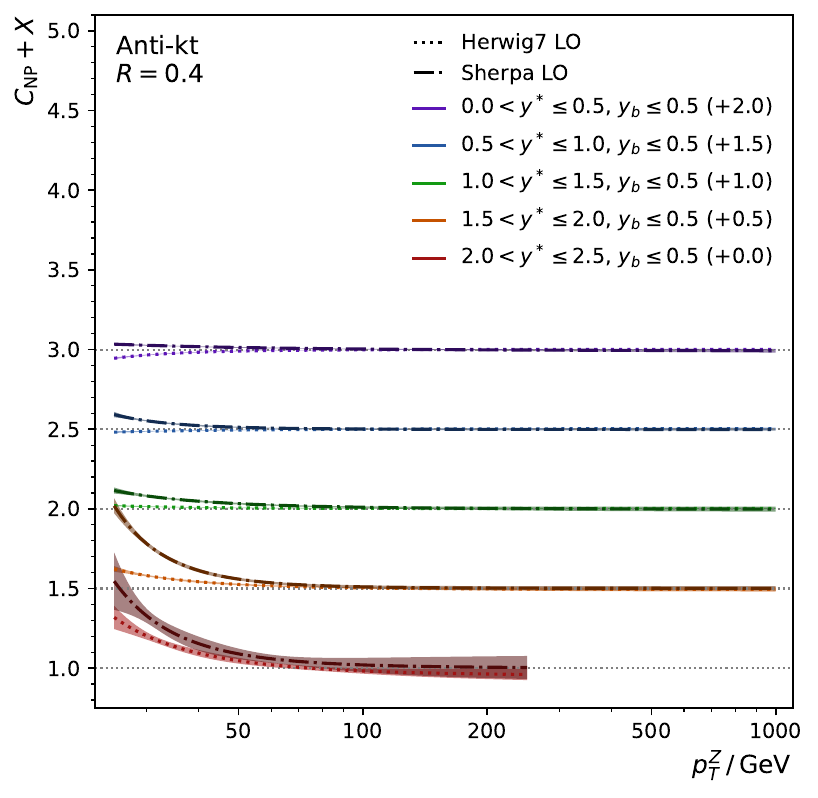}\hfill%
  \includegraphics[width=0.49\linewidth]{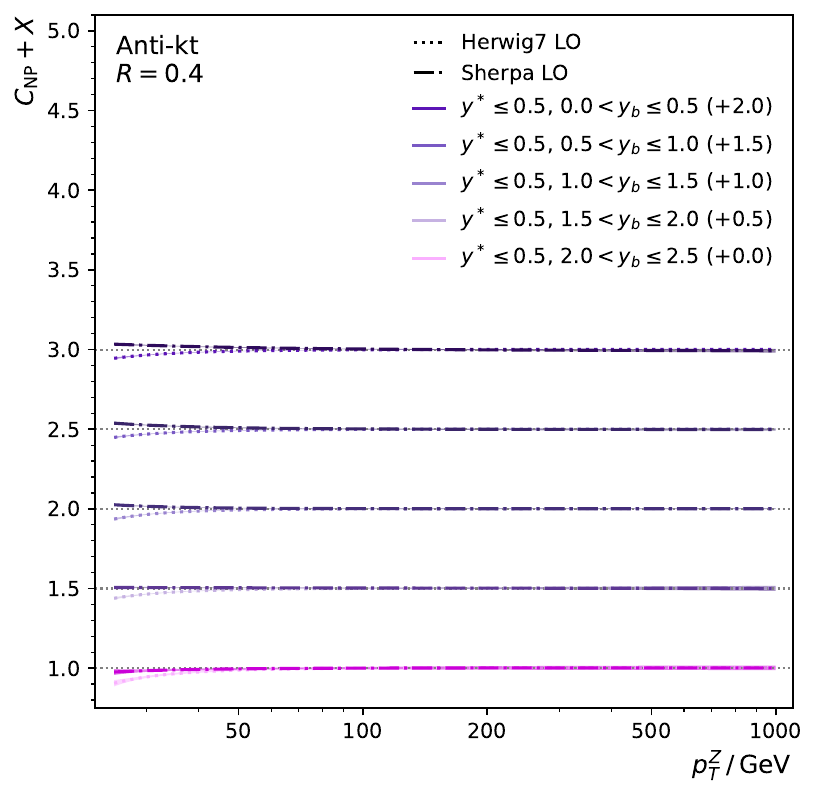}\\
  \includegraphics[width=0.49\linewidth]{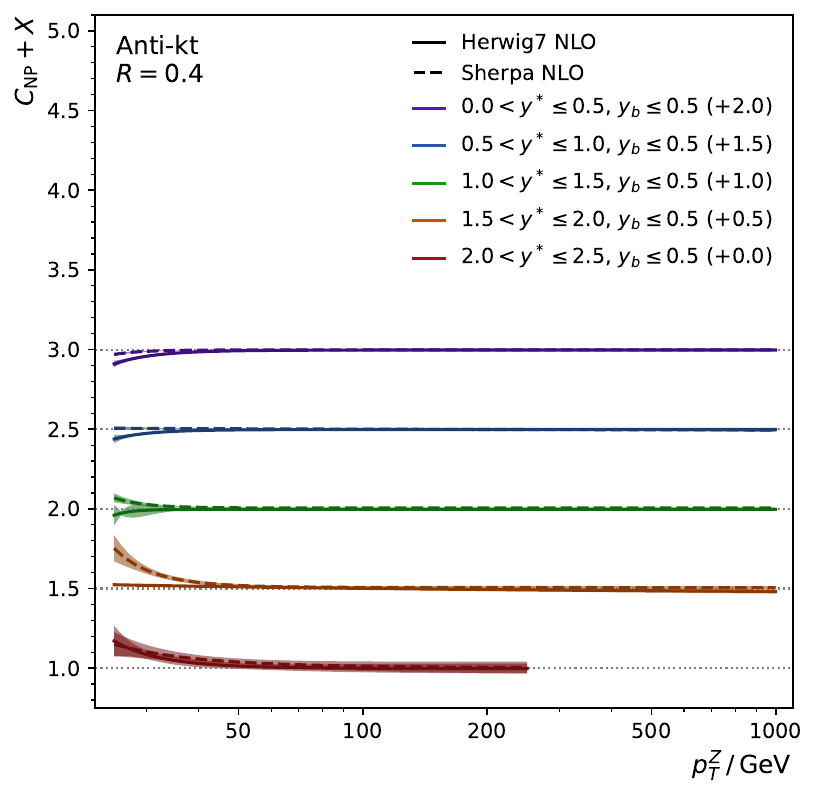}\hfill%
  \includegraphics[width=0.49\linewidth]{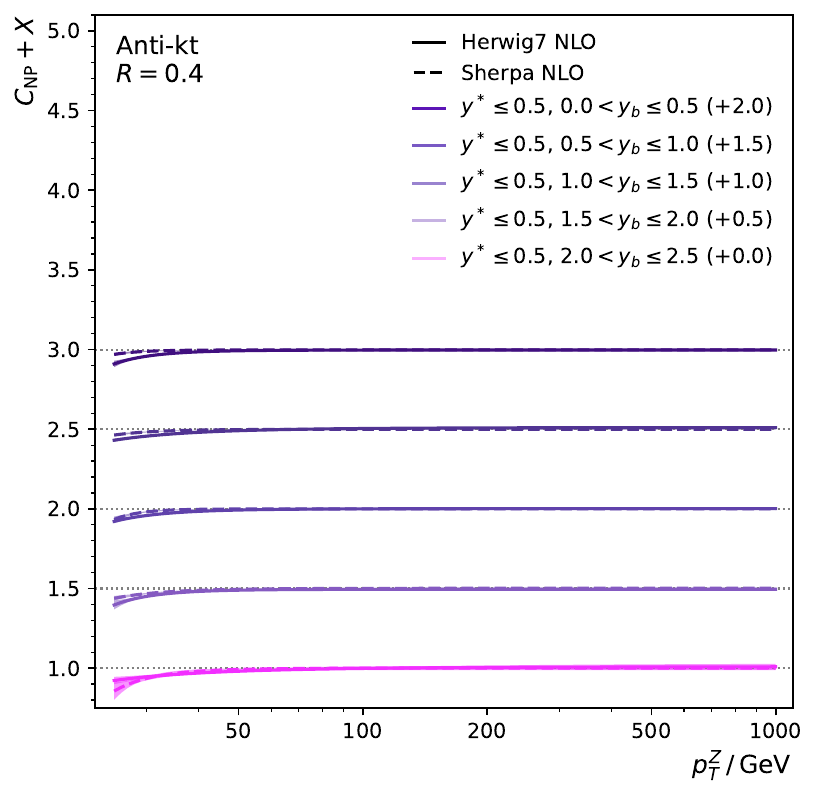}
  \caption{Comparison of \cnp vs.\ \ptz between \herwig (dotted \& solid lines) and \sherpa predictions (dash-dotted \& dashed lines) using LO (top
    row) and NLO MEs (bottom row) for anti-\kt jets with $R=0.4$. Bands indicate the statistical uncertainty. The fitted curves are shown for a series
    of phase space intervals increasing in \ystar for $\yboost < 0.5$ (left) and increasing in \yboost for $\ystar < 0.5$ (right). The factors are
    shifted with respect to unity by constant offsets.}
  \label{fig:zjetAK4-np-herwig-sherpa-lo}
\end{figure*}

As a cross-check, the NP correction factors derived using \herwig are compared in Fig.~\ref{fig:zjetAK4-np-herwig-sherpa-lo} to
\sherpa~\cite{Gleisberg:2008ta,Sherpa:2019gpd} at LO and NLO accuracy. The NP corrections derived with \sherpa are overall larger than the ones
predicted by \herwig. The general features, however, of smaller corrections at NLO than at LO and the \ystar dependence are reproduced by \sherpa.

\begin{figure*}
  \centering%
  \includegraphics[width=0.49\linewidth]{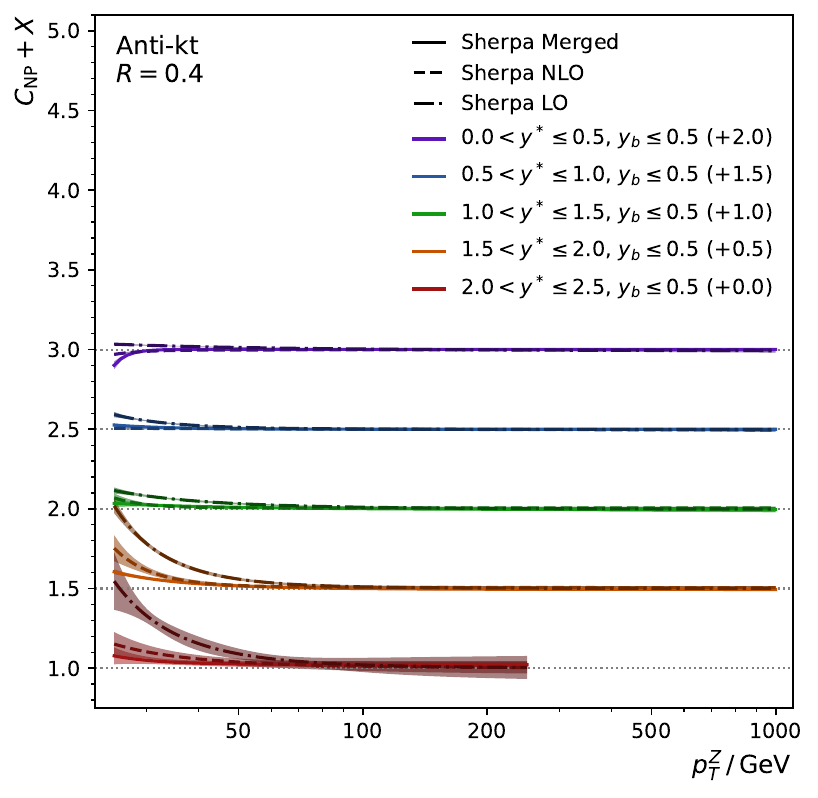}\hfill%
  \includegraphics[width=0.49\linewidth]{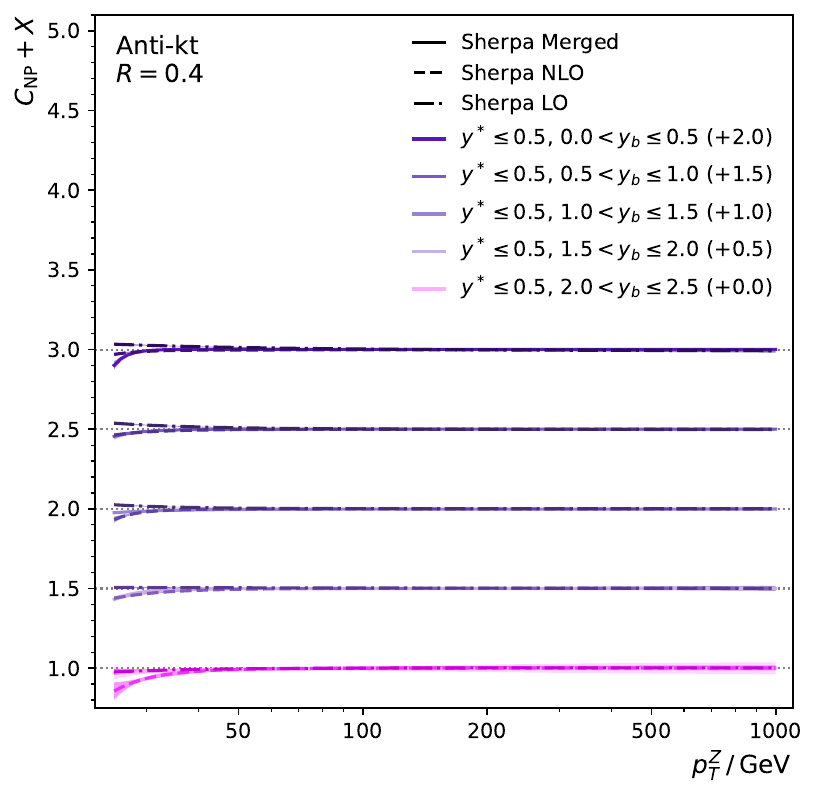}
  \caption{Comparison of \cnp vs.\ \ptz among \sherpa predictions using LO (dash-dotted), NLO (dashed), and multi-leg merged MEs (solid lines) for
    anti-\kt jets with $R=0.4$. Bands indicate the statistical uncertainty. The fitted curves are shown for a series of phase space intervals
    increasing in \ystar for $\yboost < 0.5$ (left) and increasing in \yboost for $\ystar < 0.5$ (right). The factors are shifted with respect to
    unity by constant offsets.}
  \label{fig:zjetAK4-np-sherpa-lo-nlo-merged}
\end{figure*}

To verify the trend observed with increasing parton multiplicity at ME level, NP correction factors with multi-leg merged Z+jet with one, two, and
three jets at NLO ME accuracy are generated using \sherpa.  The corresponding NP correction factors are shown in
Fig.~\ref{fig:zjetAK4-np-sherpa-lo-nlo-merged}.  Again a similar dependence on \ystar is visible in the NP correction factors derived from the
multi-leg merged Z+jet event sample.  Also, the behaviour of overall smaller NP corrections with higher multiplicity in the ME for the event
generation is confirmed. Consistently, no dependence on \yboost is observed.

This dependence of the NP correction factors on \ystar and on the perturbative order of the MEs in generating the hard amplitudes indicates that part
of the observed NP correction factors might, in fact, originate in the perturbative modelling. The labelling as ``nonperturbative'' might therefore
not be completely appropriate. Interestingly, smaller NP corrections with NLO+PS MC event generators as compared to LO+PS and rapidity-dependent PS
effects have previously been reported in the context of inclusive jet production binned in jet \pt and absolute rapidity in
Refs.~\cite{Dooling:2012uw,CMS:2014qtp}. %
Furthermore, the fact that this \ystar dependence is only observed for the Z+jet but not for the dijet process motivates a more thorough
investigation.

\section{Hadronisation \& MPI effects}
\label{sec:hadmpicorr}

According to our definition the NP correction factors account for two effects, hadronisation and the UE, which we attempt to study individually in the
following.

The parton final state is determined by the parton showers that produce additional, visible activity in all hard events.  This also holds for the
additional events from MPI that are superimposed with the primary hard events.  After the parton showers, all partons in the final state are assigned
a colour line, which in turn assigns a colour partner to each parton.  This would lead to uncorrelated events from MPI and to a strong growth of
charged particle multiplicity with the number of additional hard events.  However, the observation of increasing average transverse momentum in higher
multiplicity events necessitates colour reconnection (CR) models that lead to interference between the coloured final states of the primary hard
events and the MPI events.

The modelling of colour reconnection is based on the idea of colour pre-confinement, \ie the observation that partons that are close to each other in
phase space (usually momentum space) tend to exchange soft, NP gluons in a way that their colour lines are reconnected to become partners with close
neighbours. While there are ideas to put this on a firm theoretical basis of a colour evolution model
\cite{Gieseke:2018gff,Platzer:2022jny,Hoang:2024nqi}, most modern event generators still model this effect by minimising a ``colour distance'' of
neighbouring partons \cite{Christiansen:2015yca,Christiansen:2015yqa,Gieseke:2012ft}.  For our consideration, which is purely based on observations of
the final state, we should simply keep in mind that colour reconnection also links the MPI with the hadronisation models that we took as signposts for
the NP modelling.

In order to nonetheless get an approximate, individual handle on the hadronisation and MPI effects, respectively, we compute hadronisation and MPI
correction factors similar to Eq.~\ref{eq:nppo} as
\begin{equation}
  C_{\text{Had}}^{\text{PO}} = \frac{\sigma_{\text{ME+PS+Had}}^{\text{PO}}}{\sigma_{\text{ME+PS}}^{\text{PO}}}%
  \label{eq:hadpo}
\end{equation}
and
\begin{equation}
  C_{\text{MPI}}^{\text{PO}} = \frac{\sigma_{\text{ME+PS+MPI}}^{\text{PO}}}{\sigma_{\text{ME+PS}}^{\text{PO}}} ~~.%
  \label{eq:mpipo}
\end{equation}

\begin{figure*}
  \centering%
  \includegraphics[width=0.49\linewidth]{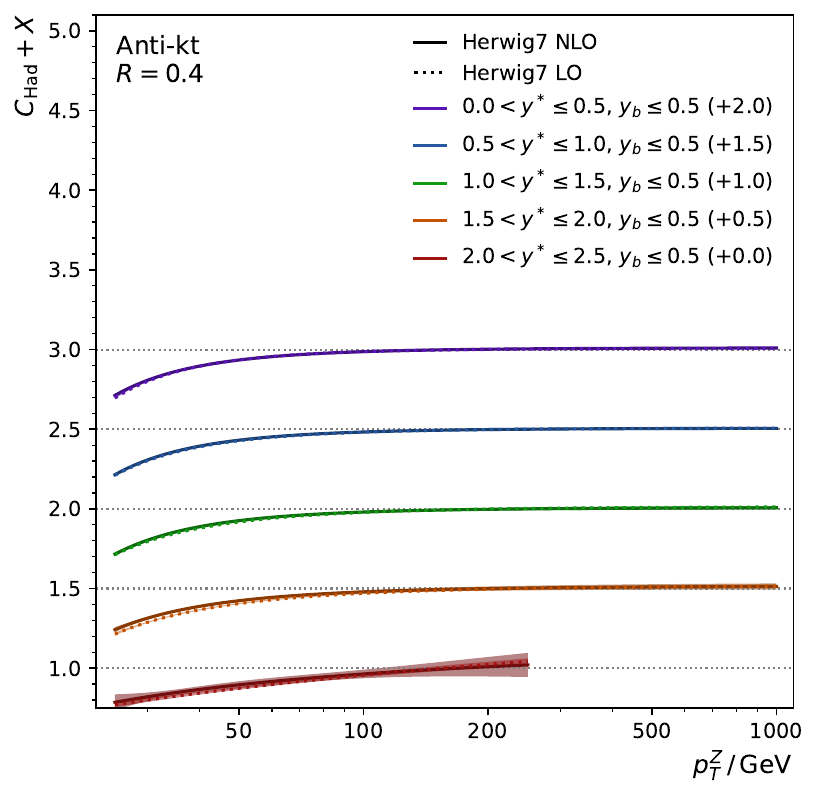}\hfill%
  \includegraphics[width=0.49\linewidth]{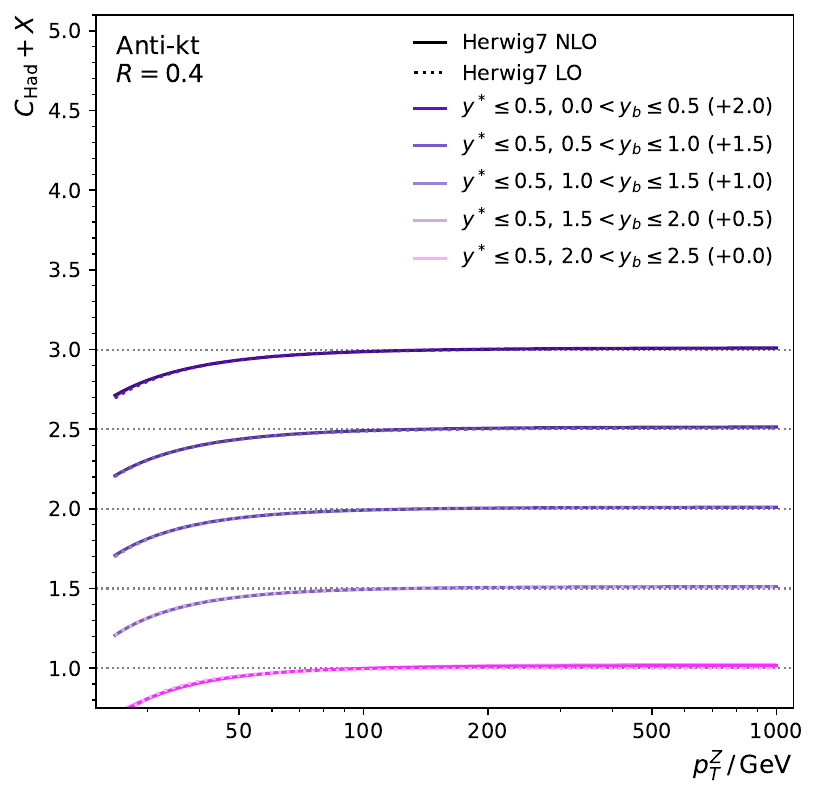}
  \caption{Same as Fig.~\ref{fig:zjetAK4-np-herwig-lo-nlo} but for hadronisation effects, \chad, only.}
  \label{fig:zjetAK4-had-herwig-lo-nlo}
\end{figure*}

The hadronisation correction factors \chad for Z+jet production with anti-\kt jets with radius parameter $R=0.4$ computed with \herwig at LO and NLO
accuracy in pQCD of the ME are shown in Fig.~\ref{fig:zjetAK4-had-herwig-lo-nlo}.  They are consistently approaching a value of unity for high
\ptz. Towards small \ptz the downwards deviation from unity increases as expected for out-of-cone effects through hadronisation.  However, they show
no significant dependence on \yboost, \ystar or the perturbative order of the generated samples.

\begin{figure*}
  \centering%
  \includegraphics[width=0.49\linewidth]{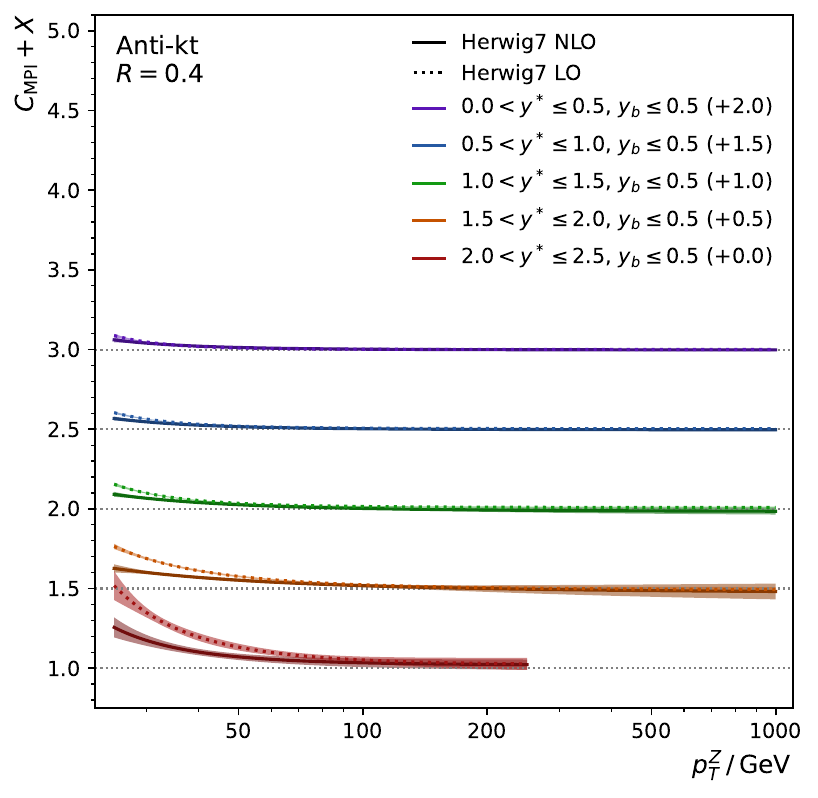}\hfill%
  \includegraphics[width=0.49\linewidth]{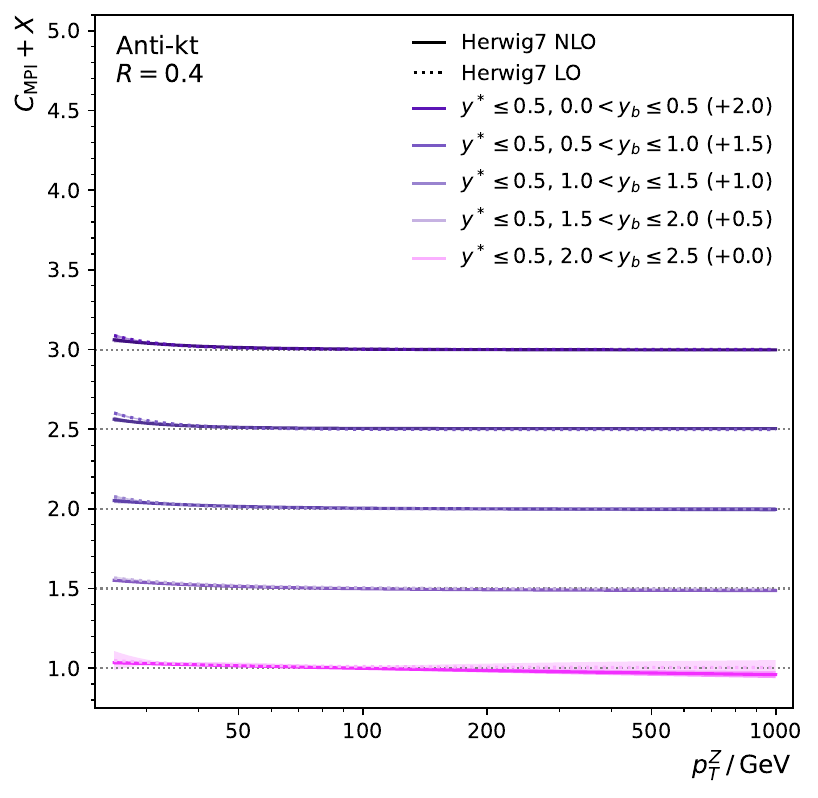}
  \caption{Same as Fig.~\ref{fig:zjetAK4-np-herwig-lo-nlo} but for MPI effects, \cmpi, only.}
  \label{fig:zjetAK4-mpi-herwig-lo-nlo}
\end{figure*}

On the contrary, the corresponding MPI correction factors \cmpi for Z+jet production computed with \herwig, shown in
Fig.~\ref{fig:zjetAK4-mpi-herwig-lo-nlo}, reveal trends versus \ystar and the perturbative order resembling the ones observed for the NP correction
factors in Sect.~\ref{sec:npcorr}.  They also approach unity towards high \ptz, while for small \ptz this time the upwards deviation from unity
increases reflecting a higher share of MPI partons migrating into a jet. The magnitude of the increase towards small \ptz correlates with \ystar and
decreases with increasing perturbative order.  These observations indicate that the origin of the observed trends for the NP correction factors lies,
at least partially, in the MPI model.

\section{Underlying Event analysis}
\label{sec:ue}

As observed in Sect.~\ref{sec:hadmpicorr}, both hadronisation and the UE each play a role for the NP corrections.  Although these two effects cannot
be completely decoupled, the trends with \ystar and the perturbative order in the NP correction factors noticed in Sect.~\ref{sec:npcorr} must mainly
originate from MPI effects.

In this section, a closer look at the NP effects, focusing especially on MPI, is presented.  For this purpose, observables of ``Rick-Field-type'' UE
analyses are utilised that are typically employed as sensitive observables in tuning and modelling studies~\cite{CDF:2001onq,CDF:2004jod,CDF:2010pdo}.

\subsection{Analysis}
\label{subsec:ueana}

\begin{figure}
  \centering%
  \includegraphics[width=0.5\linewidth]{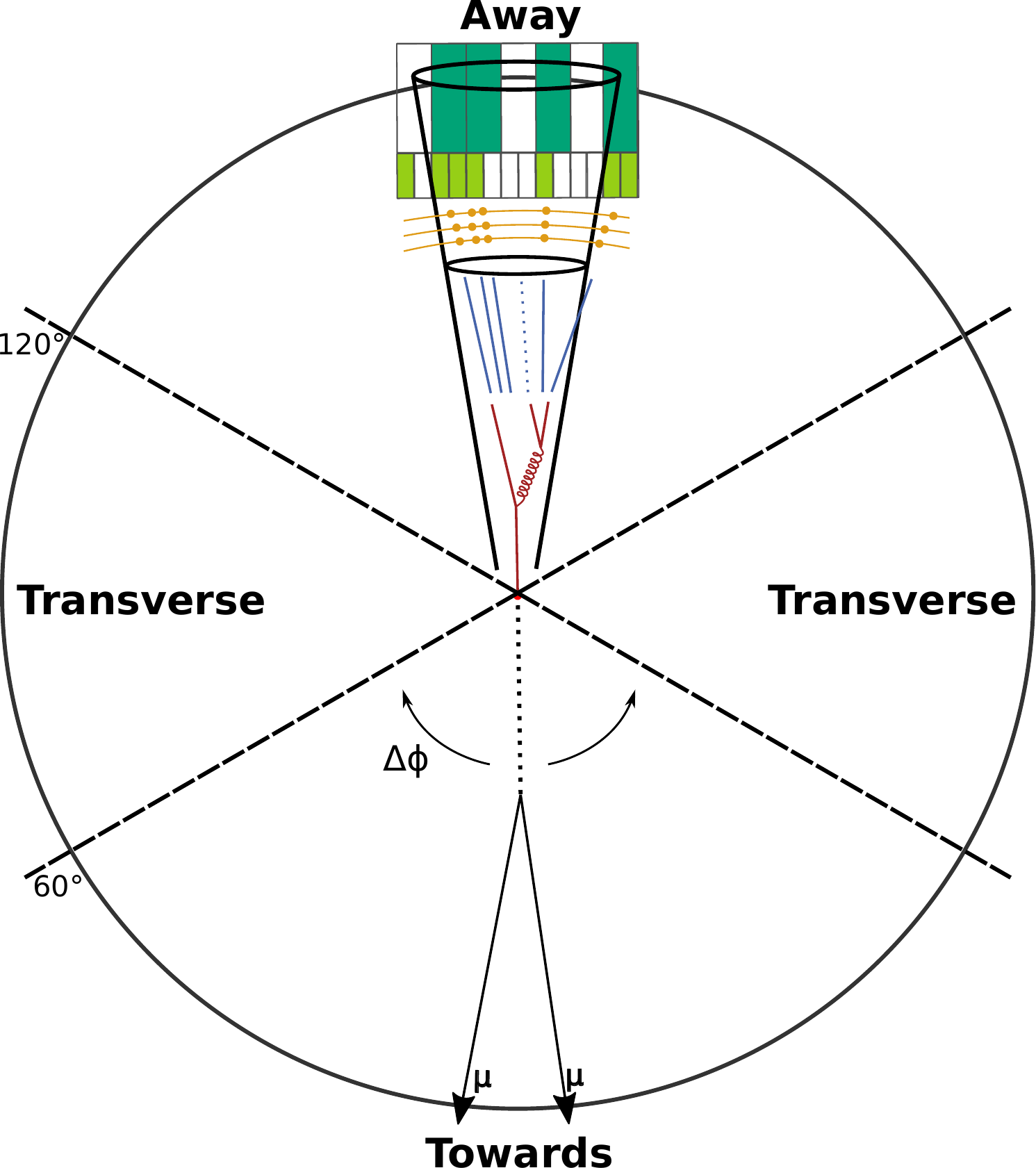}
  \caption{Visualisation of the analysed phase space for the UE analysis. The away, towards, and transverse regions are defined relative to the
    direction in azimuthal angle $\phi$ of the dimuon system.}
  \label{fig:UEPS}
\end{figure}

We pre-select Z+jet events as in the previous sections in bins of \yboost and \ystar, however, with two modifications for the UE analysis: First,
there is no minimum \ptz required, and secondly only charged particles are considered for the UE observables.  Moreover, the \pt of each such particle
must exceed 500\MeV. Then, the momentum of the lepton pair, \ie the Z~boson, is chosen to define the direction of the leading particle and henceforth
the origin of the azimuthal angle $\phi$ for the UE analysis.  All other charged particles, including the ones from the jet, are categorised as
belonging to one of three azimuthal regions, depending on the azimuthal angular difference $\Delta\phi$ relative to the leading particle, such that
$-\pi < \Delta\phi \leq \pi$.  Particles with $|\Delta\phi| \leq \pi/3$ are in the \emph{towards} region, moving roughly forward relative to the
leading particle.  Particles moving in the direction of the recoiling jet, defined by $|\Delta\phi| > 2\pi/3$, are assigned to the \emph{away} region.
The remaining region with $\pi/3 < |\Delta\phi| \leq 2\pi/3$ is called the \emph{transverse} region. Figure~\ref{fig:UEPS} illustrates the three
regions relative to the (downwards-going) leading particle and the recoiling jet.

In a dijet or minimum bias event, the transverse region allows us to have a closer look at those particles that are furthest away from the leading,
hard interaction.  These particles are associated with the UE activity, as it is observed that, beyond a certain threshold, the activity in the UE is
more or less decorrelated from the hard interaction.  In order to quantify the activity of the UE, two different observables are considered in bins of
different ``hardness'' of the leading particle characterised by its transverse momentum. For a given leading-particle transverse momentum, the average
number of charged particles $N_\mathrm{ch}$ or the sum of absolute transverse momenta $\sum p_T^{\rm ch}$ in either of the previously defined regions
are considered as a measure for the activity in these regions.

In dijet or minimum bias events, the activity in the towards and away region is strongly associated with the transverse momentum of the leading
particle.  Hence, both observables tend to grow with increasing hardness of the event.  The parton radiation associated with the hardness of the event
will mostly populate the towards and away region, but can spill over into the transverse region.  If there is a portion of the event that is more or
less uncorrelated to the hard event, as it is modelled in most MPI models, the activity in the transverse region should be dominated by the MPI
activity or at least be the region that is least overshadowed from particles associated with the hard event.

In events with the Z~boson as the leading particle, this is somewhat different.  The hardness of the event is associated with the transverse momentum
of the Z~boson.  With increasing transverse momentum of the Z~boson the recoiling activity will mostly be in the away region, while the region around
the lepton pair, \ie the towards region, should be the one with the least activity from the UE{}.

In this work, it is of interest whether the large NP corrections for relatively low transverse momentum of the Z~boson, particularly for increasing
values of \ystar, are really related to the activity from MPI. In these events, where the Z~boson and the recoiling jet are getting closer and closer
to the beam axis, there might also be a strong influence from parton radiation around the beam direction. Therefore, the UE observables are studied
with different generators and for different matching and merging scenarios in order to discriminate activity due to hard radiation in the event from
MPI activity.

\subsection{Observations}
\label{subsec:ueobs}

To illustrate our findings, plots of the activity $\sum p_T^{\rm ch}$ in the three different regions are shown.  The results for $N_{\rm ch}$ are
qualitatively similar and therefore omitted in this discussion.  These observables are derived from events generated with \herwig and \sherpa with
both LO and NLO accuracy of the hard Z+jet process. In addition, results obtained from the multileg-merged samples from \sherpa are compared.

The description of UE observables in inclusive Z+jet events with \herwig has been considered before and is known to give a reasonable agreement with
data~\cite{CDF:2010pdo,CMS:2012oqb,ATLAS:2014yqy,CMS:2017ngy,ATLAS:2019ocl}.  We therefore focus on the UE observables in differential \ybys bins as
defined in Sec.~\ref{sec:phasespace}.  In the following discussions, the outermost \ystar bin ($2.0 \leq \ystar < 2.5$) is omitted, because it is
subject to large statistical fluctuations.

\begin{figure*}
  \centering%
  \includegraphics[width=0.49\linewidth]{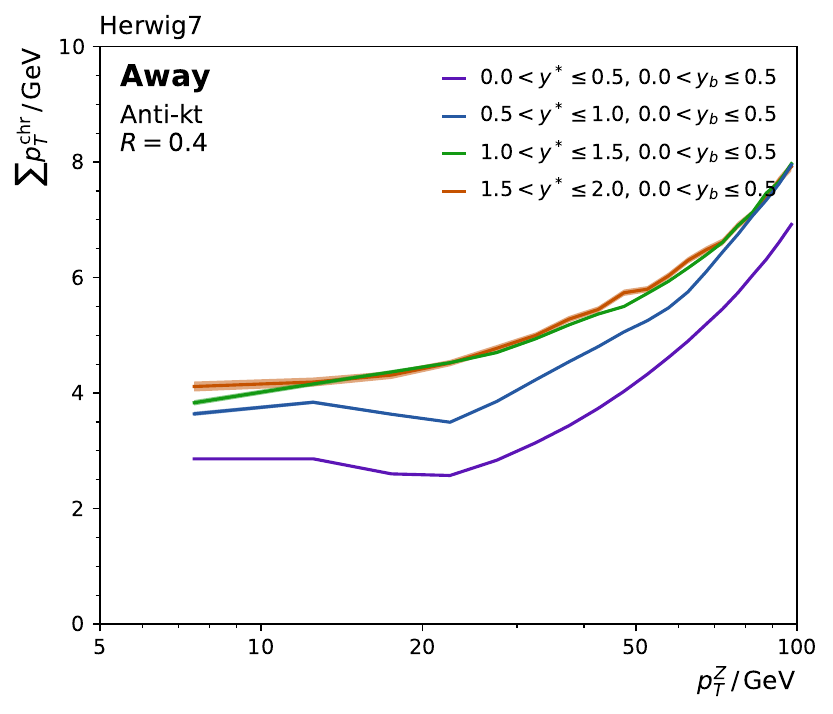}\hfill%
  \includegraphics[width=0.49\linewidth]{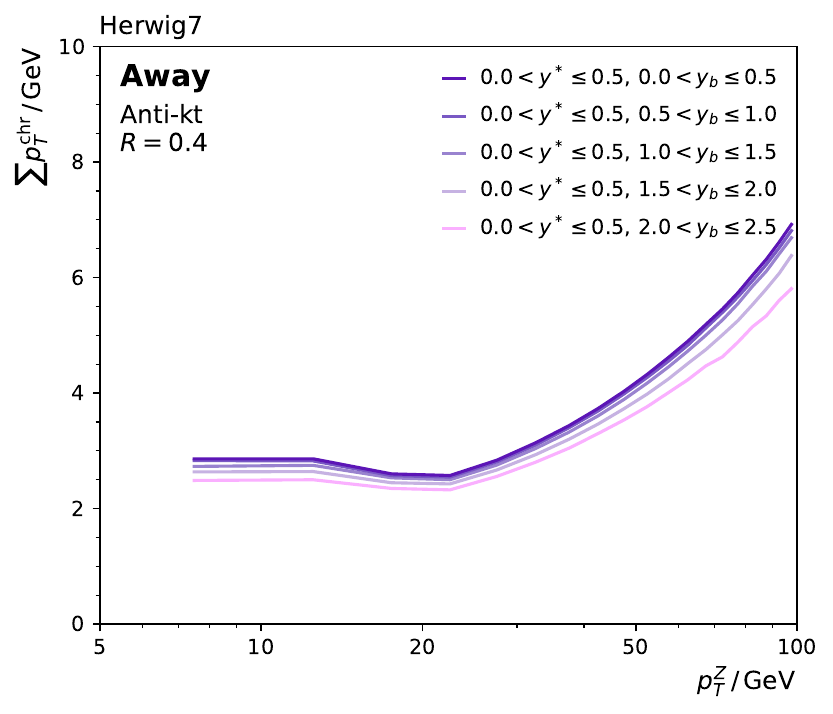}\\
  \includegraphics[width=0.49\linewidth]{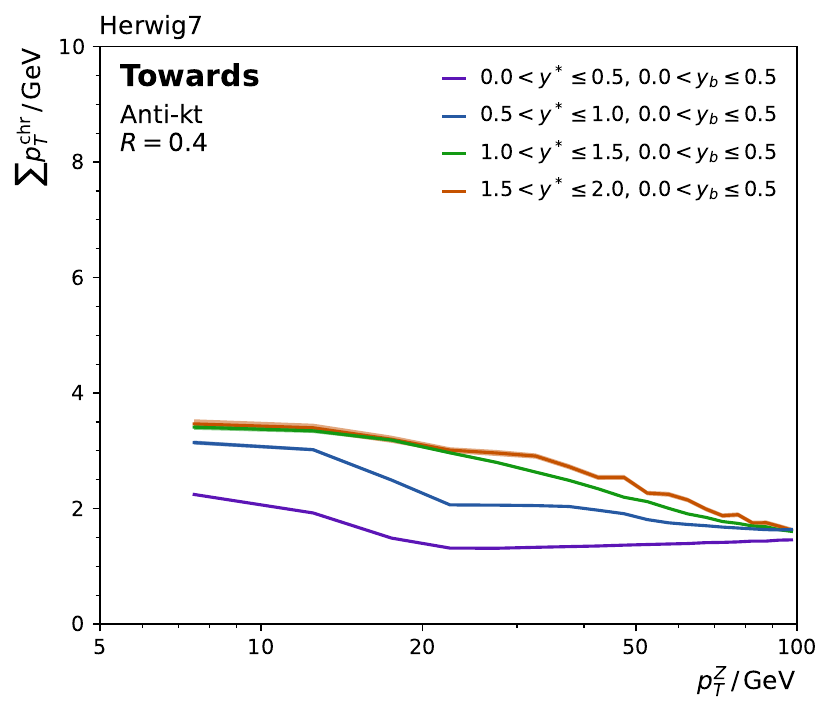}\hfill%
  \includegraphics[width=0.49\linewidth]{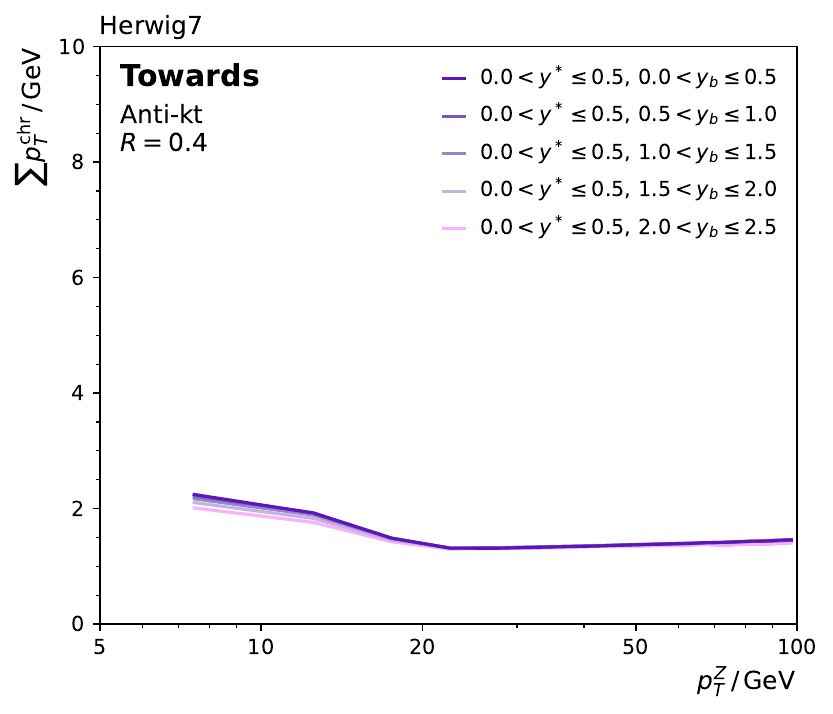}\\
  \includegraphics[width=0.49\linewidth]{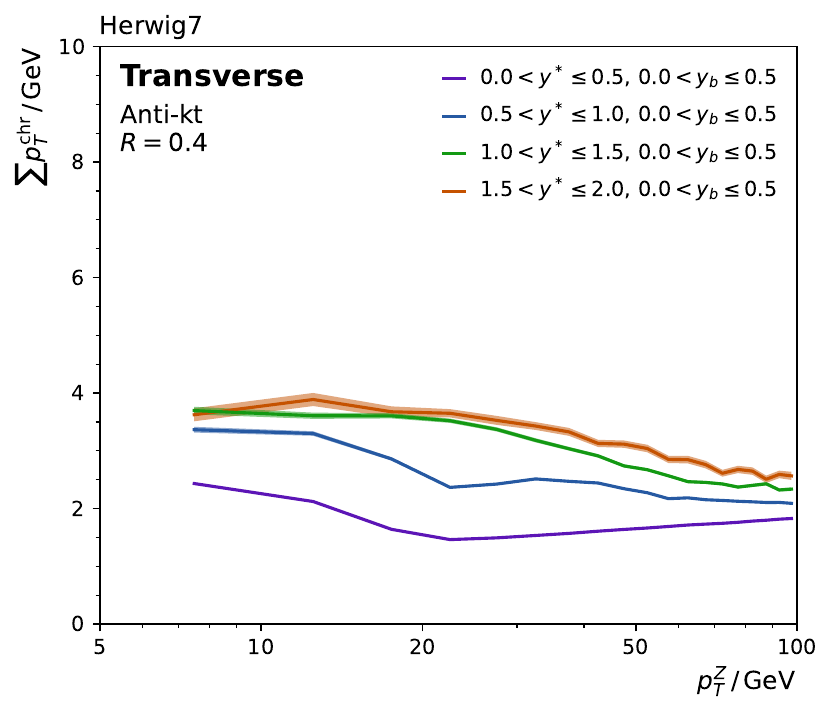}\hfill%
  \includegraphics[width=0.49\linewidth]{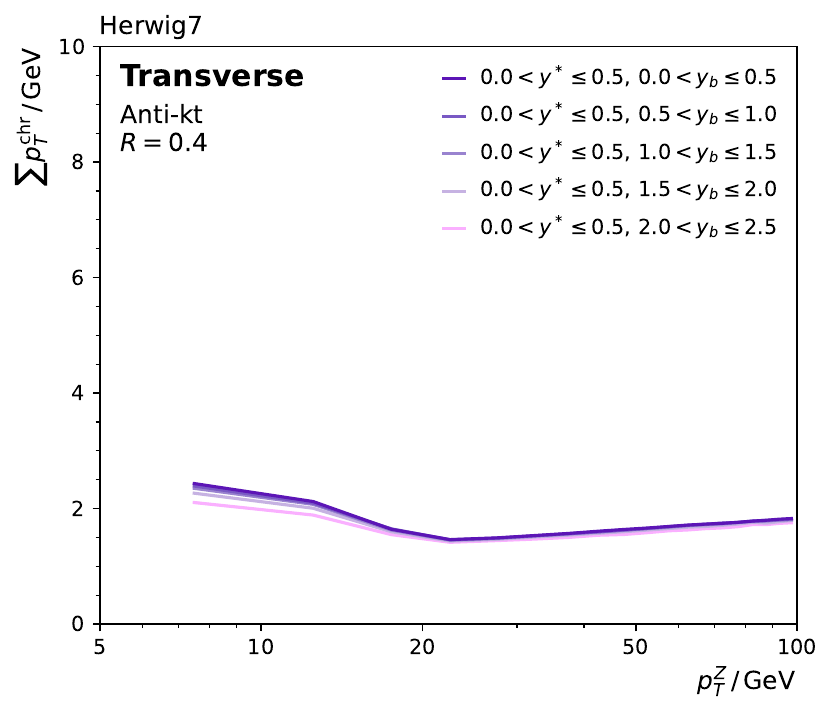}
  \caption{Comparison of the sum of transverse momenta of all charged particles, $\sum p_T^\mathrm{chr}$, in bins of \ptz, \yboost, and \ystar for the
    away (top), towards (middle), and transverse region (bottom row) as predicted by \herwig using LO MEs for event generation up to particle level.
    The values are shown for a series of phase space intervals increasing in \ystar for $\yboost < 0.5$ (left) and increasing in \yboost for
    $\ystar < 0.5$ (right).}
  \label{fig:UEherwigLOpTsum}
\end{figure*}

\begin{figure}
  \centering%
  \begin{overpic}[width=0.5\linewidth,tics=10]{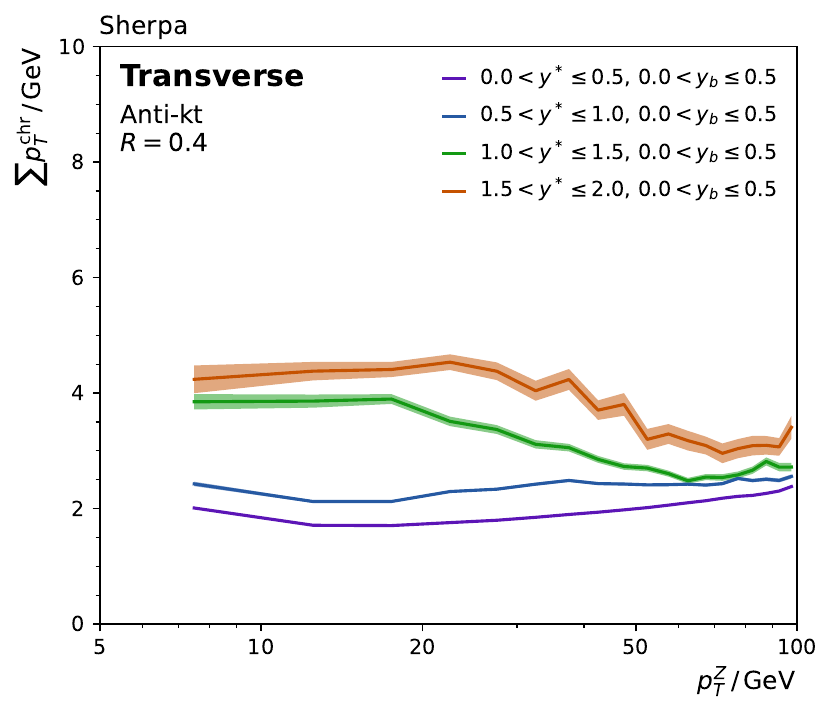}
    \put (24,81.7) {\sffamily\tiny LO}
  \end{overpic}\\
  \begin{overpic}[width=0.5\linewidth,tics=10]{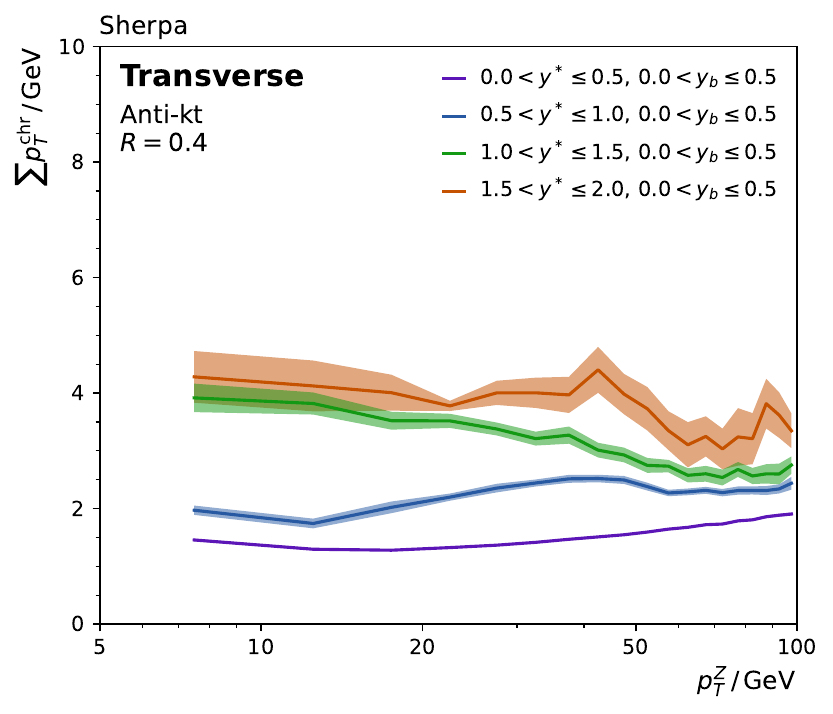}
    \put (24,81.7) {\sffamily\tiny NLO}
  \end{overpic}\\
  \begin{overpic}[width=0.5\linewidth,tics=10]{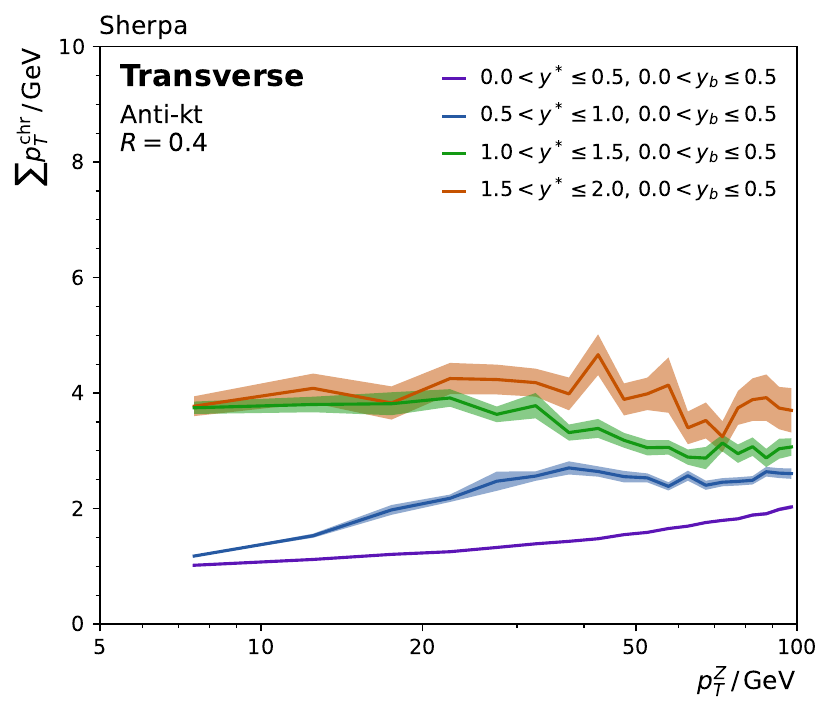}
    \put (24,81.7) {\sffamily\tiny Merged}
  \end{overpic}
  \caption{Comparison of the sum of transverse momenta of all charged particles, $\sum p_T^\mathrm{chr}$, for central \yboost in bins of \ptz and
    \ystar for the transverse region as predicted by \sherpa using LO (top), NLO (middle), and multileg-merged MEs (bottom) for event generation up to
    particle level.}
  \label{fig:UEsherpaOrderChainpTsumTransverseFull}
\end{figure}

\begin{figure}
  \centering%
  \begin{overpic}[width=0.5\linewidth,tics=10]{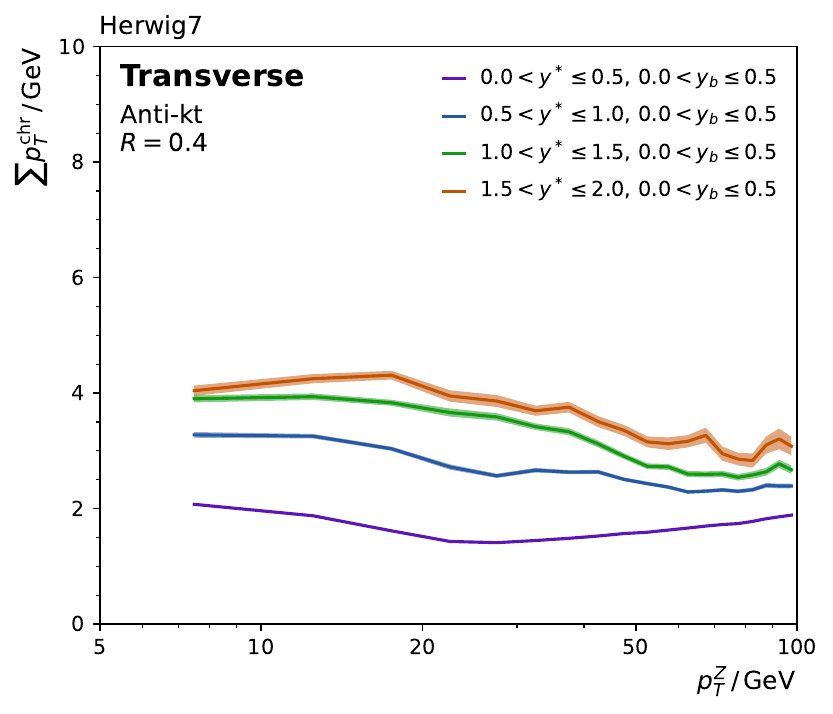}
    \put (27,81.7) {\sffamily\tiny ME+PS+Had+MPI}
  \end{overpic}\\
  \begin{overpic}[width=0.5\linewidth,tics=10]{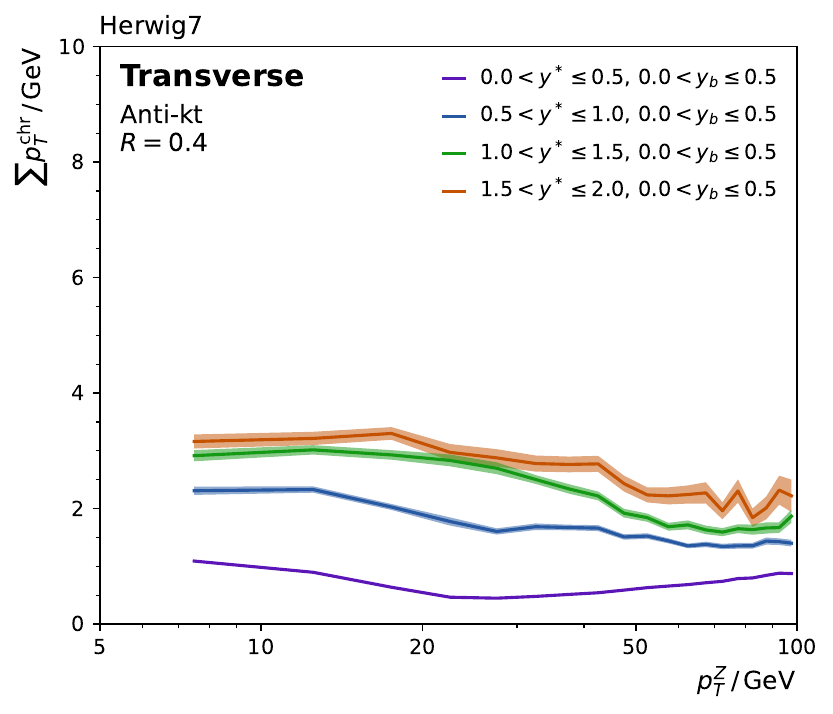}
    \put (27,81.7) {\sffamily\tiny ME+PS+Had}
  \end{overpic}\\
  \begin{overpic}[width=0.5\linewidth,tics=10]{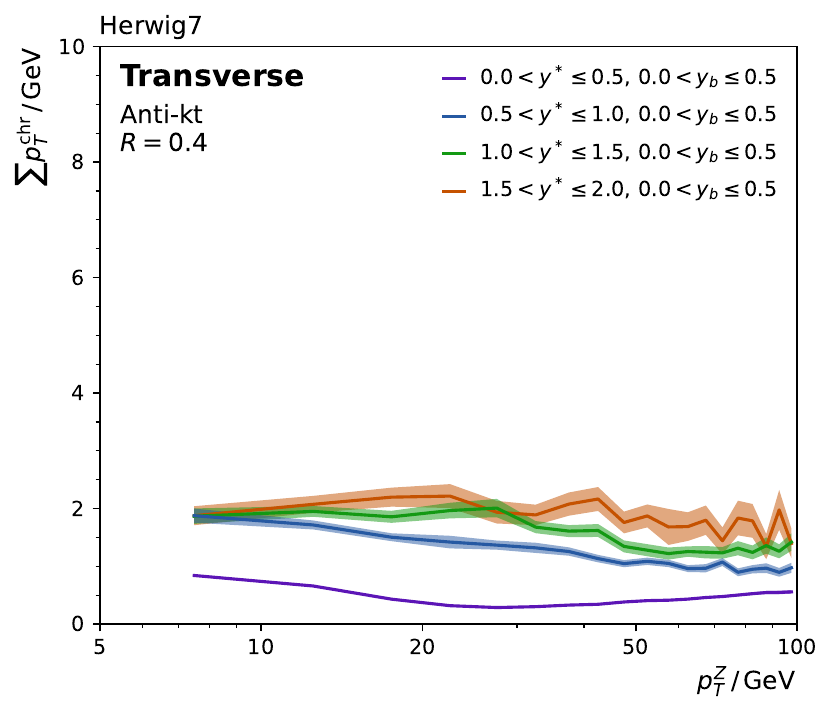}
    \put (27,81.7) {\sffamily\tiny ME+PS}
  \end{overpic}
  \caption{Comparison of the sum of transverse momenta of all charged particles, $\sum p_T^\mathrm{chr}$, for central \yboost in bins of \ptz and
    \ystar for the transverse region as predicted by \herwig using NLO MEs for event generation up to particle level with MPI (top), without MPI
    (middle), and up to parton level only, \ie hadronisation and MPI off (bottom).}
  \label{fig:UEherwigNLOpTsumTransverseTriggerChain}
\end{figure}

In Fig.~\ref{fig:UEherwigLOpTsum} from top to bottom the results from LO ME event generation for the activity in the three regions away, towards and
transverse are shown.  In the left column \yboost is fixed to the central bin ($\yboost<0.5$), while \ystar is varied.  In the right column the
respective results with \ystar fixed, while \yboost varies, are shown.  It is apparent that the UE activity varies strongly with \ystar while there is
no apparent variation with \yboost{}.  In the simplified back-to-back picture this means that the activity increases only when the central
back-to-back hard event is tilted with \ystar but not when a central back-to-back hard event is boosted along the beam axis.  The na\"ive expectation
is that any contamination of activity from the beam or MPIs should occur in both situations likewise.  Instead, the increased activity with \ystar in
all regions hints at some relation between the hard event topology and the observed UE activity.

Considering the away region, it is found that the activity here is strongly correlated with the transverse momentum of the Z~boson and increases with
\ystar.  This reflects mostly the activity in the recoiling jet, which for increasing \ystar pulls in more particles from the beam direction.  Also, we
can clearly see that, as expected, the activity in this region increases significantly from around 20\,GeV, which is the selection cut for the jet
itself. Below 20\,\GeV in \ptz the \pt of the ``balancing'' jet must be compensated by other hadronic activity in the towards region. In fact, the
hadronic activity in the towards and transverse regions decrease slightly from $\ptz = 7.5\GeV$ up to 20\GeV{}.  Increasing \yboost at central \ystar
in Fig.~\ref{fig:UEherwigLOpTsum} (right column) does not exhibit significant changes in activity in any region.

Two cross-checks are performed. First, \herwig was used with a hard process at NLO instead of LO accuracy (not shown). Generally, the results are
similar to the LO case. In particular, only a dependence on \ystar is observed but not on \yboost. The activity in the transverse region is slightly
lower at NLO than at LO\@.  Secondly, \sherpa was employed with the ME at LO, NLO, and with multi-leg merging as shown for the transverse region in
Fig.~\ref{fig:UEsherpaOrderChainpTsumTransverseFull}. As with \herwig the UE activity in the transverse region strongly depends on \ystar but not on
\yboost, and decreases slightly with a more accurate simulation of the hard event.

Finally, in Fig.~\ref{fig:UEherwigNLOpTsumTransverseTriggerChain} the same NLO simulation with \herwig with the full generation up to particle level
is compared to two partial generation chains. More specifically, first, only the MPI is turned off (ME+PS+Had), and secondly both the MPI and
hadronisation (ME+PS) are switched off. In comparison to the full generation mostly the overall normalisation changes suggesting the effect, that the MPI essentially
adds activity to the event, but remains broadly uncorrelated with the hard event, as does hadronisation. In summary, hadronisation and MPI alone
cannot fully explain the \ystar dependence and perturbative effects significantly contribute as well.

\section{Summary and outlook}
\label{sec:summary}

We find that nonperturbative (NP) corrections for Z+jet events in the triple-differential measurement of the cross section strongly depend on the
scattering angle of the hard process in the centre-of-mass frame, \ystar, but barely on the boost of the hard scatter, \yboost{}.  The event
generation with hadronisation effects switched off hinted at the Underlying Event (UE) simulation as being responsible for this extra activity.
However, looking into this effect with the help of a typical UE analysis in the same triple-differential binning has shown that a substantial part of
the extra activity must be of perturbative origin. The UE analysis shows the same strong dependence on \ystar, even when the multiple parton
interactions (MPI) as a model for the UE are switched off.  Albeit this UE observable has not yet been measured triple-differentially, we find
consistency between the simulation with the event generators \herwig and \sherpa.  In contrast, the NP correction factors for the dijet case do not
exhibit this strong dependence on \ystar.

In summary, we find that NP corrections, derived in the conventional way as defined in Eq.~\ref{eq:nppo}, may not be entirely of NP origin. Moreover,
they are not universal among different hard processes. The inclusion of unobserved jets will lead to additional activity that should not be attributed
to non-perturbative effects but alter the correction factors by a possibly large amount. In our specific case this means that in order to simulate the
corrections for Z+jets one should include a number of additional jets via multijet merging, possibly at NLO, in order to obtain appropriate correction
factors. Furthermore, to clarify the observation of the \ystar dependent UE activity in Z+jet event, we recommend measuring the UE activity
triple-differentially.

\section*{Acknowledgements}
We thank Simon Plätzer for fruitful discussions during early stages of this work.


\paragraph{Funding information}
This work was supported by the German Federal Ministry of Education and Research (project funding numbers 05H2021, 05H21GUCC2, 05H21VKCCA).
The authors also acknowledge support by the state of Baden-Württemberg through bwHPC and the German Research Foundation (DFG) through grant no INST 39/963-1 FUGG
(bwForCluster NEMO).

\begin{appendix}
\numberwithin{equation}{section}

\section{Generator configurations}
\label{sec:genconfig}

In the following, a summary of the generation settings for \herwig and \sherpa are given that have been used to generate the samples used in this
work.  Their complete configurations can be found in the utilized steering files published in the \texttt{inputfiles} directory of the software
framework \mcrun~\cite{horzela_2024_14530596}.

\subsection{\herwig}
\label{sec:genconfigherwig}
An overview of the generator settings for all samples generated with \herwig is given in Table~\ref{tab:herwiggensettings}.  The samples in the table
are labelled according to the underlying process and the perturbative accuracy in QCD they are generated with.  For reference, the respective
tree-level process with the orders of the QCD and electro-weak coupling constants are given.

At the lowest final state multiplicity, the dijet samples contain events with at least two partons at ME level, the Z+jet events contain an oppositely
charged muon pair and one jet.  Further settings for the matrix element generation are the lower transverse momentum cutoff for the coloured final
state particles labelled as $k_T^\text{jet,min}$, the choice of the renormalization and factorization scales, and the chosen PDF sets.  Both
scales are set for the dijet case to the transverse momentum of the hardest parton $\max(p_T^{\text{jet}})$ and for the Z+jet case to the invariant
mass of the oppositely charged muon pair $m_{\mu^+\mu^-}$.  All are interfaced to the angular ordered parton shower in
\herwig~\cite{Bahr:2008pv,Bellm:2015jjp} and the NLO samples are matched to the parton shower by \herwig's own implementation of the MC@NLO matching
algorithm~\cite{Frixione:2002ik,Bahr:2008pv,Bellm:2015jjp}.

The dijet samples are generated with the non-perturbative models tuned to the CH3 set of parameters \cite{CMS:2020dqt} based on the default tune that
comes predefined with \herwig.  The Z+jet samples are generated without any changes to the MPI model parameters compared to the default tune that
comes with \herwig version~7.2.

\subsection{\sherpa}
\label{sec:genconfigsherpa}
For the samples generated with \sherpa an overview of the generator settings is given in Table~\ref{tab:sherpagensettings}.  They are labelled
following the same logic and the overall information given in the table and its structure follows what is given in the \herwig case described in
Appendix~\ref{sec:genconfigherwig}.

At ME level the LO and NLO Z+jet samples contain events with a oppositely charged dimuon pair plus one jet at the lowest multiplicity.  The
factorization and renormalization scales are dynamically determined with the \texttt{SCALES METS} scheme.  In all cases the generated events are
interfaced to the CSS parton shower~\cite{Schumann:2007mg}.  The NLO real corrections at the ME level in the NLO sample are matched to the parton
shower using \sherpa's implementation of the MC@NLO algorithm~\cite{Hoeche:2011fd}.  In the Merged case contributions from MEs with 0, 1, 2, and 3
partons are merged and interfaced with the parton shower using the MEPS@NLO~\cite{Hoeche:2012yf,Gehrmann:2012yg}, for zero and one, and the
MENLOPS~\cite{Hamilton:2010wh,Hoche:2010kg} techniques, for two and three partons, with a merging cut $Q_\text{cut} = 15\,\text{GeV}$.  In the LO and
NLO generations, the merging cut $Q_\text{cut} = 15\,\text{GeV}$ serves as the regulating cut for the parton emission.

The default tunes of the non-perturbative models shipped with \sherpa version 2.2.15 have been utilized in the generation of all samples.

\begin{landscape}
  \begin{table}[t]
    \centering
    \begin{tabular}{c|c|c|c|c|c|c|c|c}
      \toprule
      Process Label & tree-level MEs & $\mathcal{O}(\alps), \mathcal{O}(\alpha)$ & $k_T^{\text{jet,min}}$ & scales & PDF set & PS & Matching/Merging & Tune\\
      \midrule
      dijet LO    & $p+p \to j+j$ & 2, 0 & 20\,GeV & $\max(p_T^{\text{jet}})$ & NNPDF31\_nnlo\_as\_0118 & AO & --     & CH3 \\
      dijet NLO   & $p+p \to j+j$ & 2, 0 & 20\,GeV & $\max(p_T^{\text{jet}})$ & NNPDF31\_nnlo\_as\_0118 & AO & MC@NLO & CH3 \\
      Z+jet LO    & $p+p \to \mu^+ + \mu^- +j$ & 1, 2 & 20\,GeV & $m_{\mu^+\mu^-}$ & CT14lo & AO & --     & Default \\
      Z+jet NLO   & $p+p \to \mu^+ + \mu^- +j$ & 1, 2 & 20\,GeV & $m_{\mu^+\mu^-}$ & CT14nlo & AO & MC@NLO & Default \\
      \bottomrule
    \end{tabular}
    \caption{Generator settings for the samples generated with \herwig version 7.2.3}
    \label{tab:herwiggensettings}
  \end{table}

\begin{table}[b]
  \centering
  \begin{tabular}{c|c|c|c|c|c|c|>{\centering\arraybackslash}p{3cm}|c}
    \toprule
    Process Label & tree-level MEs & $\mathcal{O}(\alps), \mathcal{O}(\alpha)$ & $Q_\text{cut}$ & scales & PDF set & PS & Matching/Merging & Tune\\
    \midrule
    Z+jet LO     & $p+p \to \mu^+ + \mu^- +j$ & 1, 2 & 15\,GeV & SCALES METS & CT14lo & CSS & -- & Default \\
    Z+jet NLO    & $p+p \to \mu^+ + \mu^- +j$ & 1, 2 & 15\,GeV & SCALES METS & CT14nlo & CSS & MC@NLO & Default \\
    Z+jet Merged & $p+p \to \mu^+ + \mu^- \{+j+j+j\}$ & 0, 2 & 15\,GeV & SCALES METS & CT14nlo & CSS & MEPS@NLO (0,1 jets), MENLOPS (2,3 jets) & Default \\
    \bottomrule
  \end{tabular}
  \caption{Generator settings for the samples generated with \sherpa version 2.2.15}
  \label{tab:sherpagensettings}
\end{table}
\end{landscape}

\section{Additional material}
\label{sec:addmat}

\begin{figure*}[p]
  \centering%
  \includegraphics[width=0.49\linewidth]{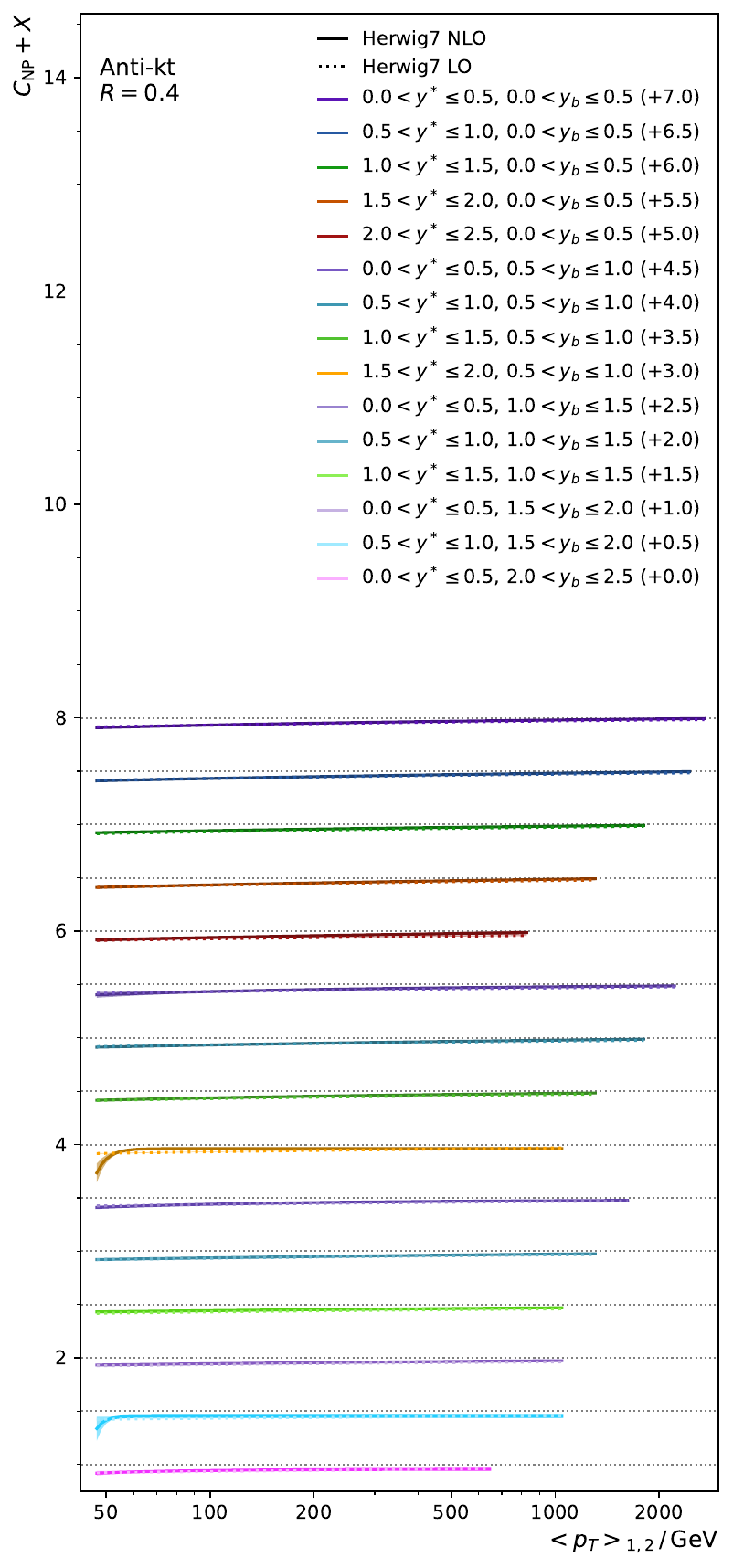}\hfill%
  \includegraphics[width=0.49\linewidth]{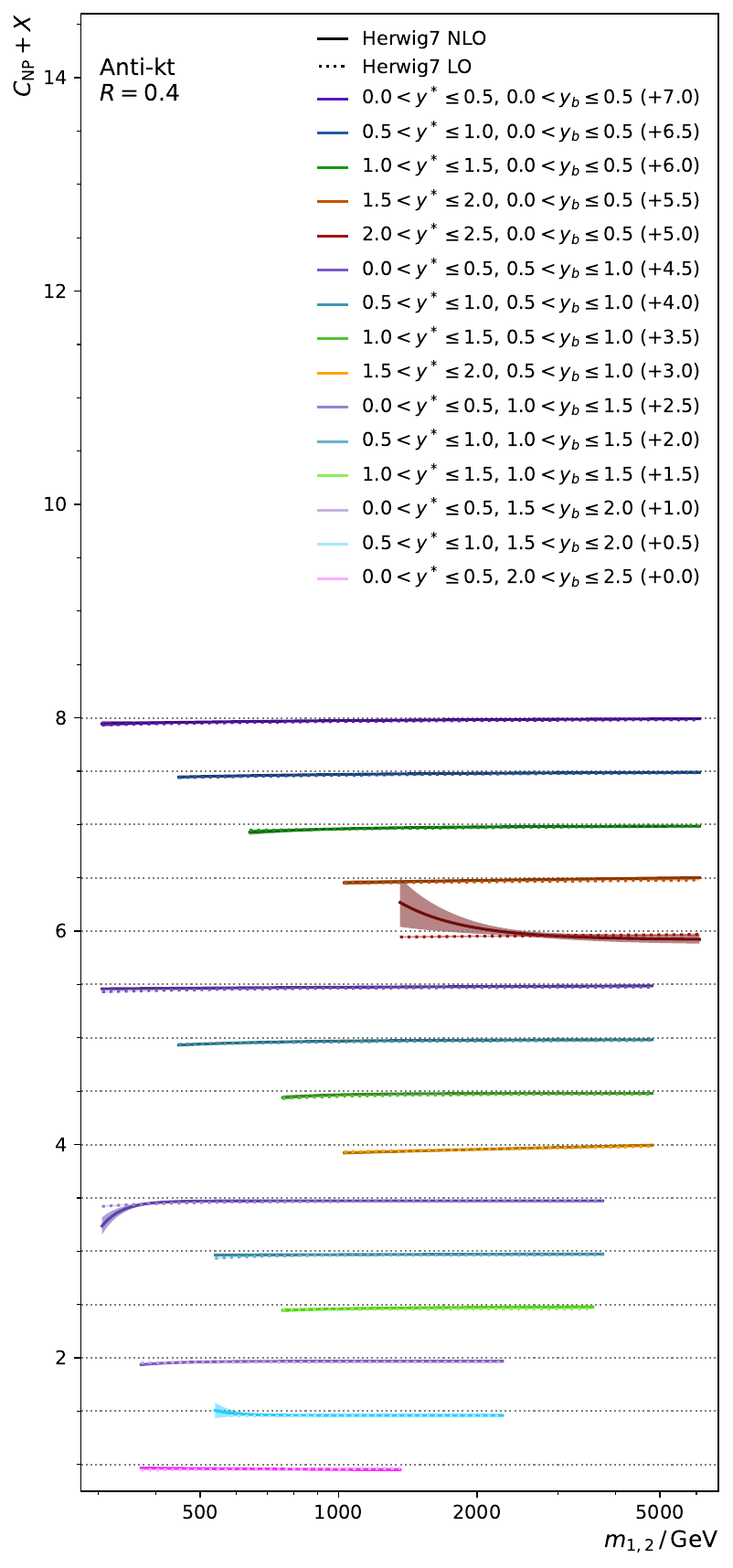}
  \caption{\cnp vs.\ dijet \ptave (left) and mass \mjj (right) using LO (dotted) and NLO MEs (solid lines) in \herwig predictions for anti-\kt jets
    with $R=0.4$.  The fitted curves are shown for all 15 defined phase space intervals in \ybys with bands indicating the statistical uncertainty.
    The factors are shifted with respect to unity by constant offsets.}
  \label{fig:dijets_herwig_ak4_all}
\end{figure*}

\begin{figure*}[p]
  \centering%
  \includegraphics[width=0.49\linewidth]{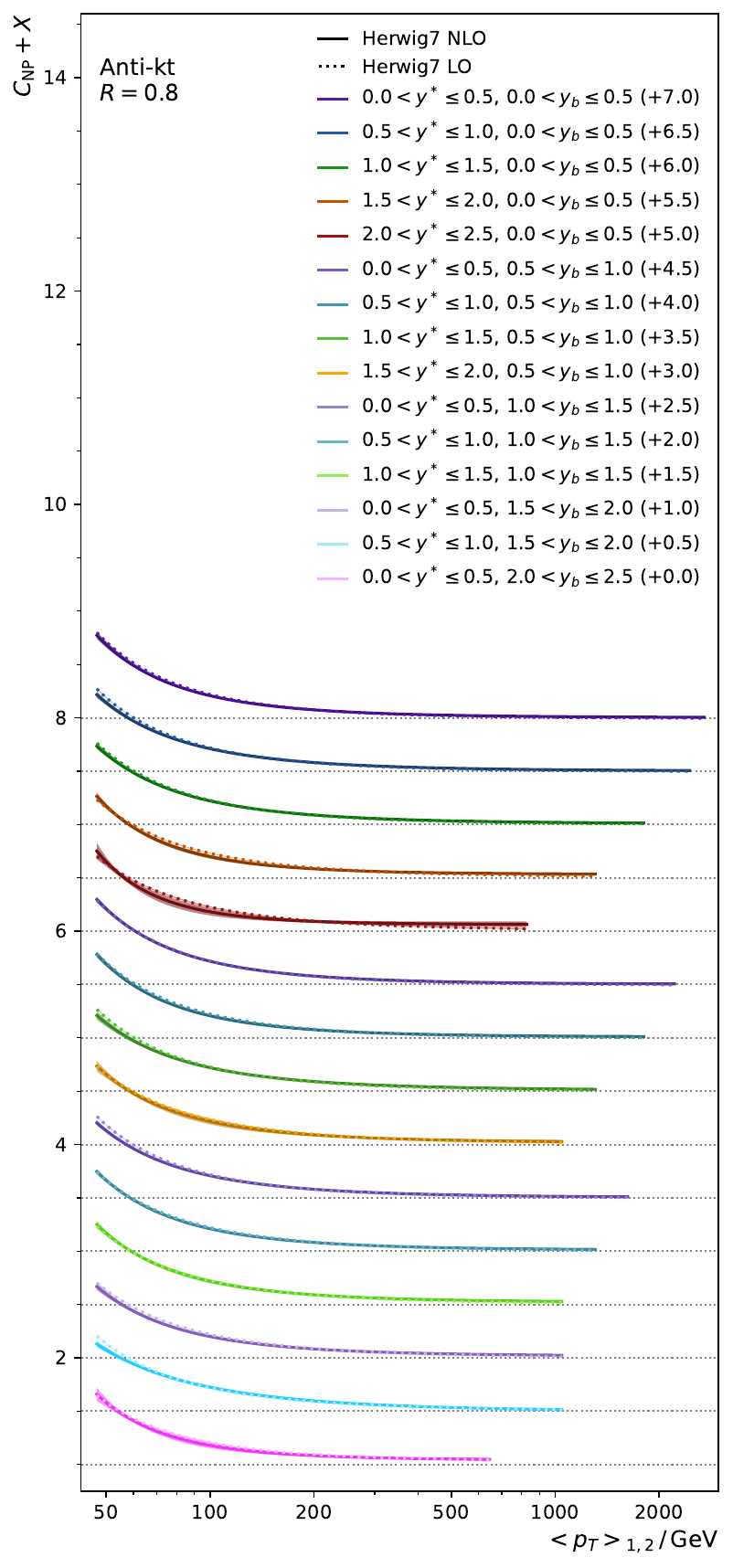}\hfill%
  \includegraphics[width=0.49\linewidth]{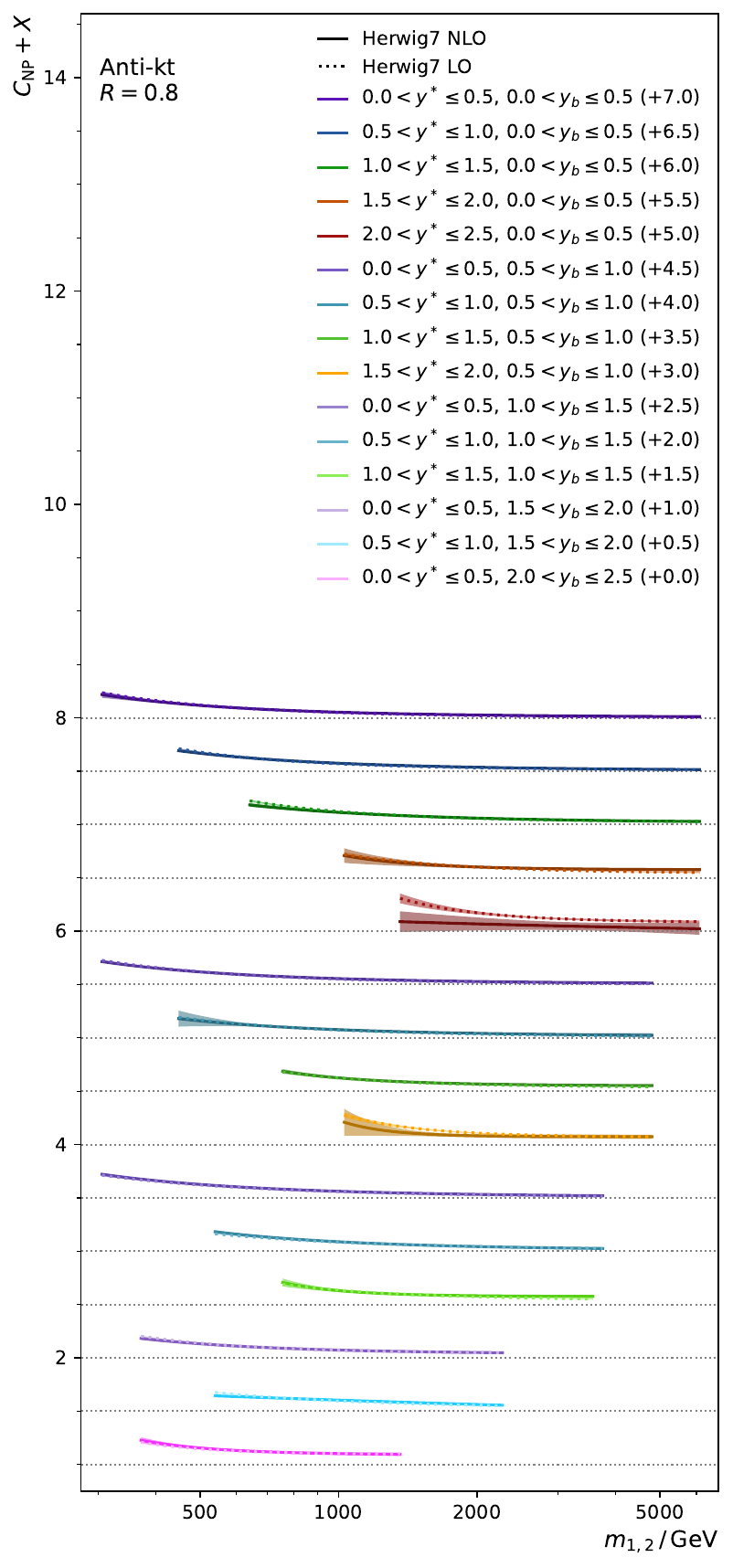}
  \caption{Same as Fig.~\ref{fig:dijets_herwig_ak4_all} but for anti-\kt jets with $R=0.8$.}
  \label{fig:dijets_herwig_ak8_all}
\end{figure*}

\begin{figure*}[p]
  \centering%
  \includegraphics[width=0.49\linewidth]{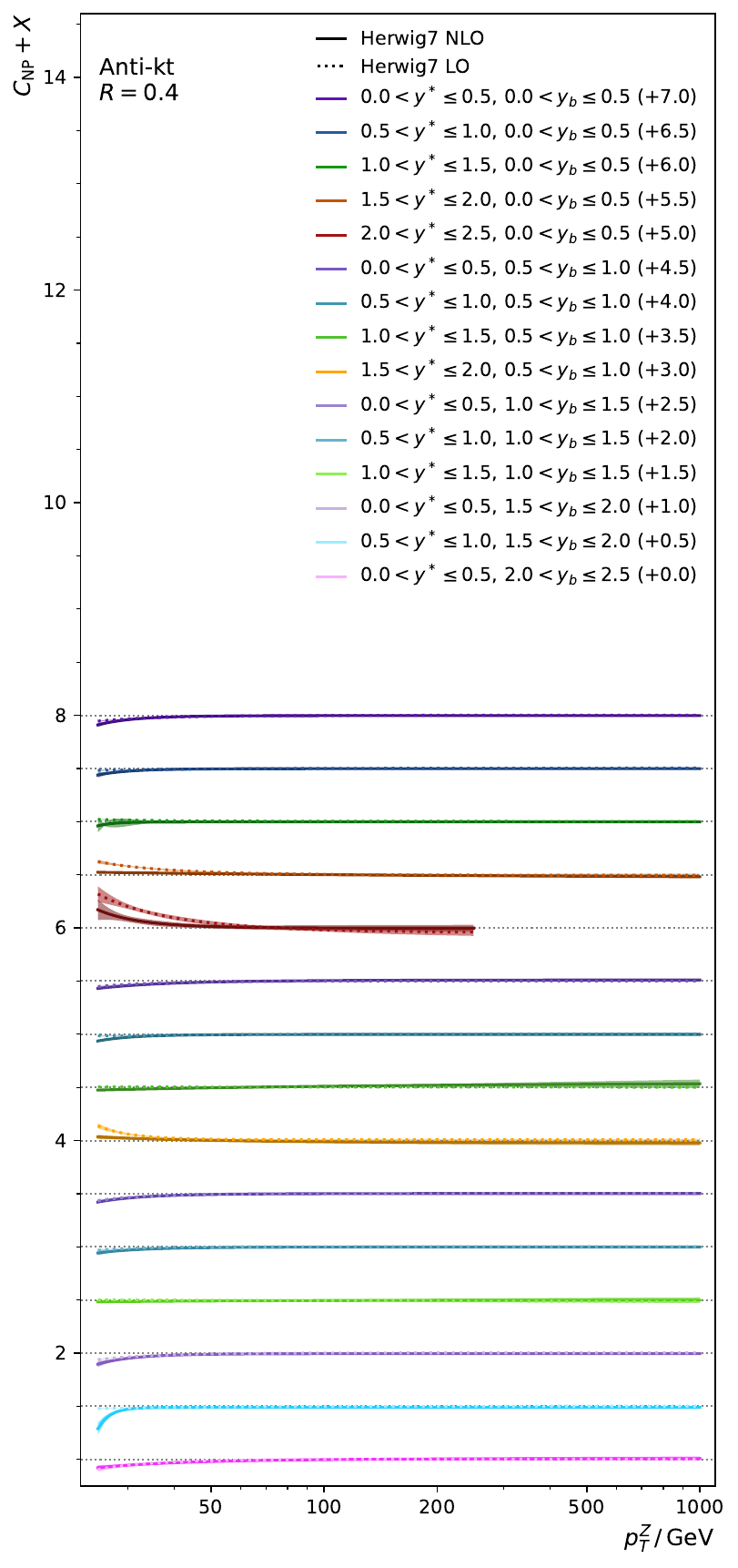}\hfill%
  \includegraphics[width=0.49\linewidth]{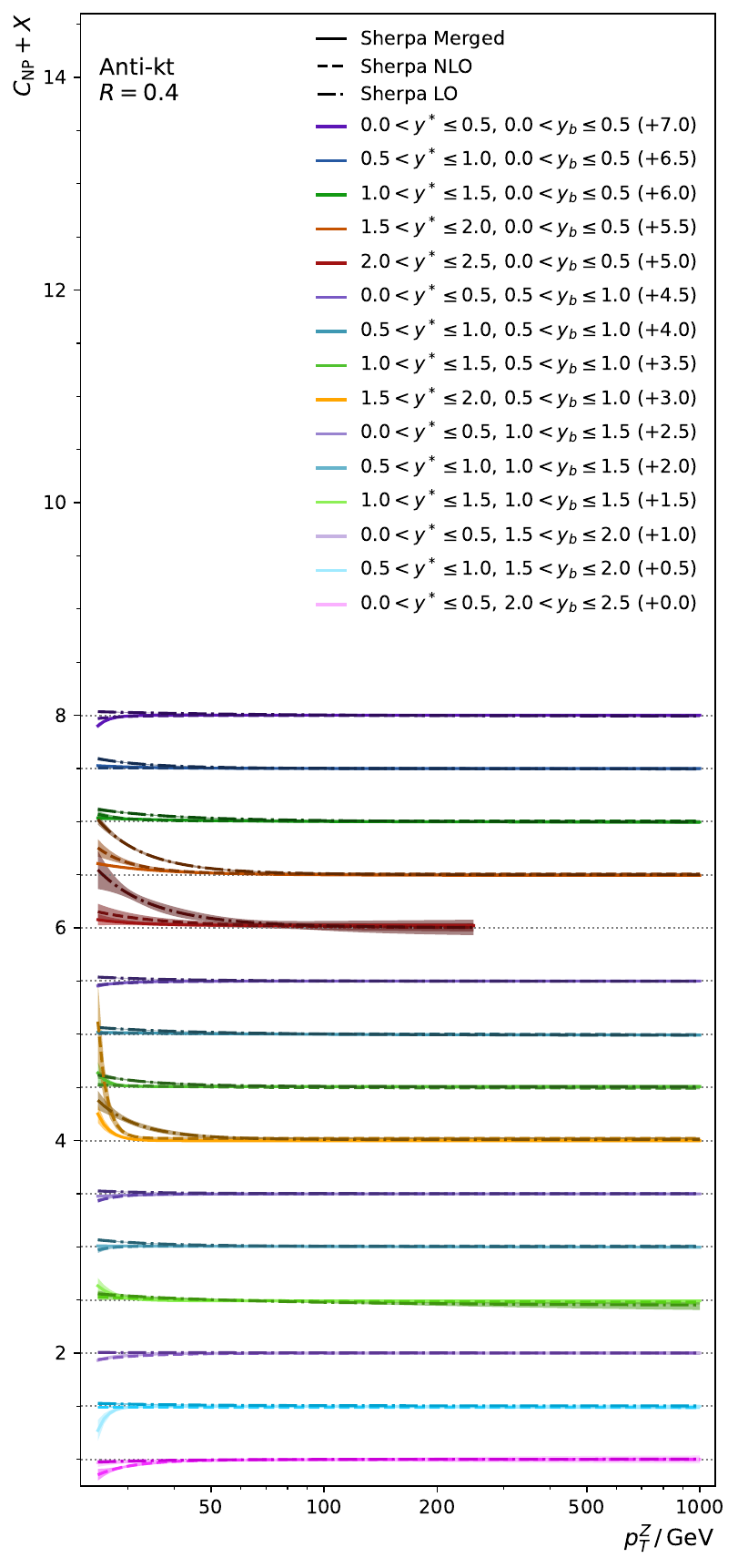}
  \caption{\cnp vs.\ \ptz from \herwig (left) using LO (dotted) and NLO MEs (solid lines), and from \sherpa (right) using LO (dash-dotted), NLO
    (dashed), and multileg-merged MEs (solid lines) for anti-\kt jets with $R=0.4$. The fitted curves are shown for all 15 defined phase space
    intervals in \ybys with bands indicating the statistical uncertainty. The factors are shifted with respect to unity by constant offsets.}
  \label{fig:zjets_ak4_zpt_all}
\end{figure*}

\begin{figure*}[p]
  \centering%
  \includegraphics[width=0.49\linewidth]{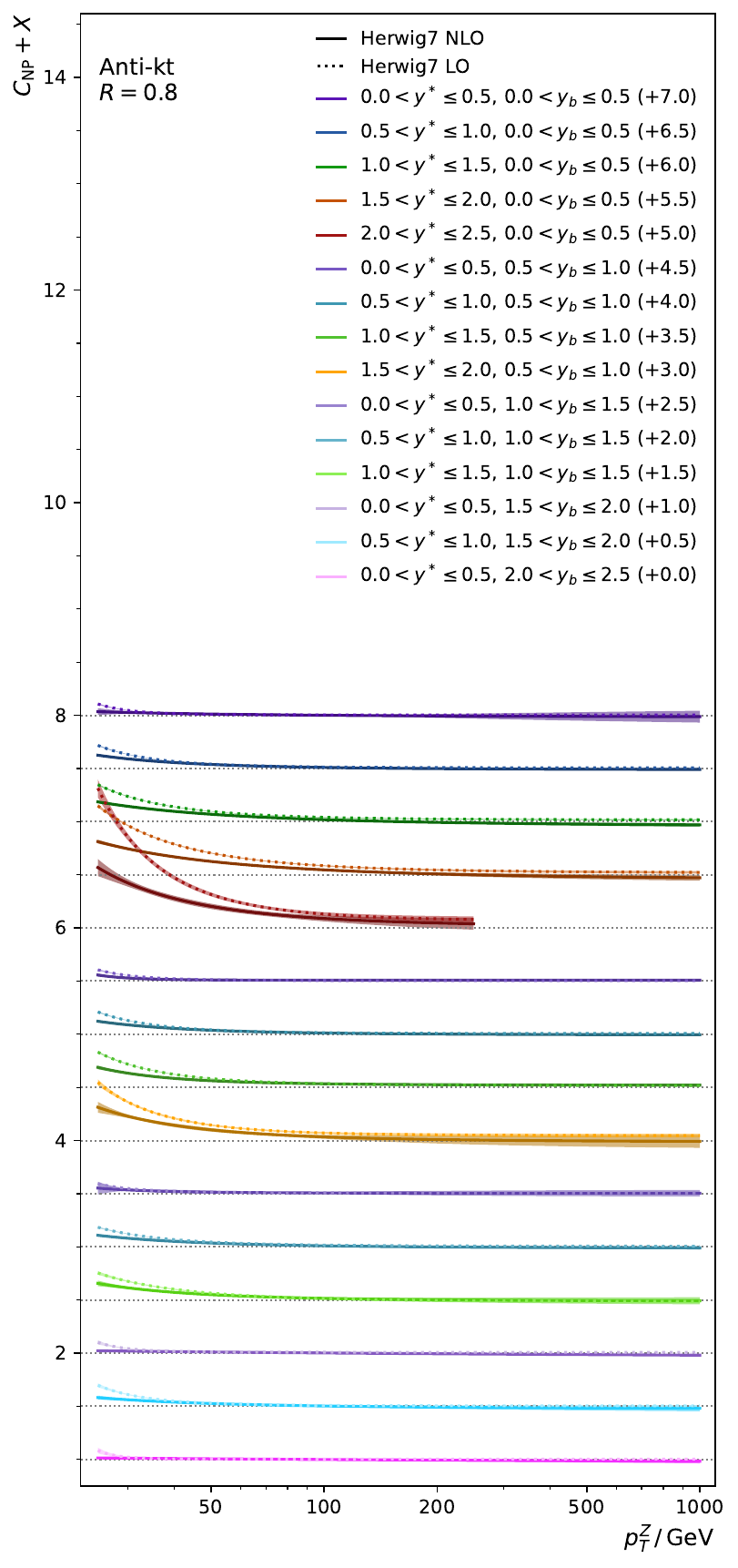}\hfill%
  \includegraphics[width=0.49\linewidth]{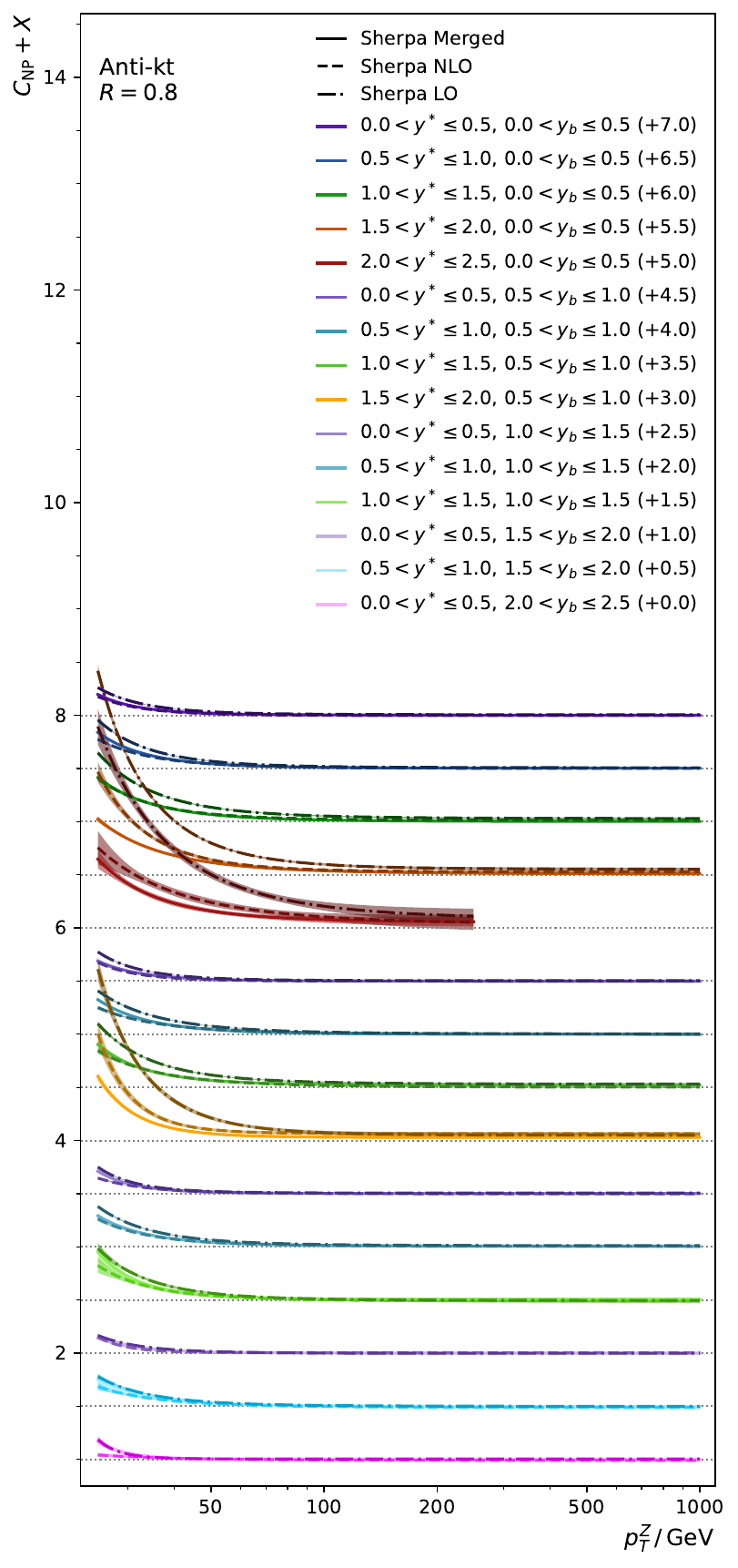}
  \caption{Same as Fig.~\ref{fig:zjets_ak4_zpt_all} but for anti-\kt jets with $R=0.8$.}
  \label{fig:zjets_ak8_zpt_all}
\end{figure*}

\begin{samepage}
In this section additional results are presented for all phase space intervals in \ybys and for the two jet size parameters $R=0.4$ and $0.8$.
Figures~\ref{fig:dijets_herwig_ak4_all} and~\ref{fig:dijets_herwig_ak8_all} show the NP~correction factors as derived from \herwig with LO and NLO MEs
for dijet production versus \ptave and \mjj. The binning scheme for dijet mass is given in Table~\ref{tab:dijetmass_binning}, which, in contrast to
\ptave, has not been extended to include results at lower \pt. %
Furthermore, Figures~\ref{fig:zjets_ak4_zpt_all} and~\ref{fig:zjets_ak8_zpt_all} show the NP correction factors as derived from \herwig and \sherpa
for all available perturbative orders of the MEs for Z+jet production versus \ptz.

Lacking a physics-related motivation for the smoothing function of the NP corrections at low scales, it should be noted that the derived curves are
valid only within the shown range and should not be extrapolated. In particular, when the inset of a steeper increase (or decrease) versus lower
values of \ptave, \mjj, or \ptz is represented by only one bin at the low edge, statistical fluctuations may lead to exaggerated slopes.

\begin{table}[hb]
  \centering%
  \caption[The binning scheme in \mjj for the dijet analysis.]{The binning scheme in \mjj for the dijet analysis. Due to limited statistical
    precision, some bins are merged in the edge bins~(E) as compared to the central bins~(C) in the plane of \ybys.}
  \begin{tabular}{cl}
    \toprule
    Binning scheme & binning labels in \ybys plane\\
    \midrule
                   & \mbox{E}\\
                   & \mbox{E\quad{}E}\\
    (cf.~Fig.~\ref{fig:ybyszj}) & \mbox{C\quad{}E\quad{}E}\\
                   & \mbox{C\quad{}C\quad{}E\quad{}E}\\
                   & \mbox{C\quad{}C\quad{}C\quad{}E\quad{}E}\\
    \midrule
                   & bin edges in $\mjj\,/\GeV$\\
    \midrule
    Central~(C) &  306, 372, 449, 539, 641, 756, 887,\\
                   & 1029, 1187, 1361, 1556, 1769, 2008,\\
                   & 2273, 2572, 2915, 3306, 3754, 4244,\\
                   & 4805, 5374, 6094 \\
    Edge~(E) & 372, 539, 756, 1029, 1361, 1769,\\
                   & 2273, 2915, 3754, 4805, 6094 \\
    \bottomrule
  \end{tabular}
  \label{tab:dijetmass_binning}
\end{table}
\end{samepage}

\end{appendix}
\clearpage





\bibliography{references.bib}

@Article{CMS:2020dqt,
  author        = {Sirunyan, Albert M and others},
  journal       = {Eur. Phys. J. C},
  title         = {{Development and validation of HERWIG 7 tunes from {CMS} underlying-event measurements}},
  year          = {2021},
  number        = {4},
  pages         = {312},
  volume        = {81},
  archiveprefix = {arXiv},
  collaboration = {CMS},
  comment       = {CH3},
  doi           = {10.1140/epjc/s10052-021-08949-5},
  eprint        = {2011.03422},
  primaryclass  = {hep-ex},
  reportnumber  = {CMS-GEN-19-001, CERN-EP-2020-182},
}

@Article{Schumann:2007mg,
  author        = {Schumann, Steffen and Krauss, Frank},
  journal       = {JHEP},
  title         = {{A Parton shower algorithm based on Catani-Seymour dipole factorisation}},
  year          = {2008},
  pages         = {038},
  volume        = {03},
  archiveprefix = {arXiv},
  doi           = {10.1088/1126-6708/2008/03/038},
  eprint        = {0709.1027},
  primaryclass  = {hep-ph},
  reportnumber  = {DCPT-07-86, IPPP-07-43},
}

@Article{Hoeche:2011fd,
  author        = {Höche, Stefan and Krauss, Frank and Schönherr, Marek and Siegert, Frank},
  journal       = {JHEP},
  title         = {{A critical appraisal of NLO+PS matching methods}},
  year          = {2012},
  pages         = {049},
  volume        = {09},
  archiveprefix = {arXiv},
  doi           = {10.1007/JHEP09(2012)049},
  eprint        = {1111.1220},
  primaryclass  = {hep-ph},
  reportnumber  = {SLAC-PUB-14661, IPPP-11-67, DCPT-11-134, LPN11-58, FR-PHENO-2011-019, MCNET-11-24},
}

@Article{Hoeche:2012yf,
  author        = {Höche, Stefan and Krauss, Frank and Schönherr, Marek and Siegert, Frank},
  journal       = {JHEP},
  title         = {{QCD matrix elements + parton showers: The NLO case}},
  year          = {2013},
  pages         = {027},
  volume        = {04},
  archiveprefix = {arXiv},
  doi           = {10.1007/JHEP04(2013)027},
  eprint        = {1207.5030},
  primaryclass  = {hep-ph},
  reportnumber  = {SLAC-PUB-15191, IPPP-12-52, DCPT-12-104, LPN12-081, MCNET-12-09, FR-PHENO-2012-017},
}

@Article{Gehrmann:2012yg,
  author        = {Gehrmann, Thomas and Höche, Stefan and Krauss, Frank and Schönherr, Marek and Siegert, Frank},
  journal       = {JHEP},
  title         = {{NLO QCD matrix elements + parton showers in $e^+e^-$ ---{\ensuremath{>}} hadrons}},
  year          = {2013},
  pages         = {144},
  volume        = {01},
  archiveprefix = {arXiv},
  doi           = {10.1007/JHEP01(2013)144},
  eprint        = {1207.5031},
  primaryclass  = {hep-ph},
  reportnumber  = {ZU-TH-11-12, SLAC-PUB-15190, IPPP-12-51, DCPT-12-102, LPN12-080, MCNET-12-08, FR-PHENO-2012-018},
}

@Article{Hamilton:2010wh,
  author        = {Hamilton, Keith and Nason, Paolo},
  journal       = {JHEP},
  title         = {{Improving NLO-parton shower matched simulations with higher order matrix elements}},
  year          = {2010},
  pages         = {039},
  volume        = {06},
  archiveprefix = {arXiv},
  doi           = {10.1007/JHEP06(2010)039},
  eprint        = {1004.1764},
  primaryclass  = {hep-ph},
  reportnumber  = {MCNET-10-05},
}

@Article{Hoche:2010kg,
  author        = {Höche, Stefan and Krauss, Frank and Schönherr, Marek and Siegert, Frank},
  journal       = {JHEP},
  title         = {{NLO matrix elements and truncated showers}},
  year          = {2011},
  pages         = {123},
  volume        = {08},
  archiveprefix = {arXiv},
  doi           = {10.1007/JHEP08(2011)123},
  eprint        = {1009.1127},
  primaryclass  = {hep-ph},
  reportnumber  = {ZU-TH-13-10, IPPP-10-73, DCPT-10-146, CERN-PH-TH-2010-195, MCNET-10-17},
}

@Article{Frixione:2002ik,
  author        = {Frixione, Stefano and Webber, Bryan R.},
  journal       = {JHEP},
  title         = {{Matching NLO QCD computations and parton shower simulations}},
  year          = {2002},
  pages         = {029},
  volume        = {06},
  archiveprefix = {arXiv},
  doi           = {10.1088/1126-6708/2002/06/029},
  eprint        = {hep-ph/0204244},
  reportnumber  = {CAVENDISH-HEP-02-01, LAPTH-905-02, GEF-TH-2-2002},
}

@Article{Dooling:2012uw,
  author        = {Dooling, S. and Gunnellini, P. and Hautmann, F. and Jung, H.},
  journal       = {Phys. Rev. D},
  title         = {Longitudinal momentum shifts, showering, and nonperturbative corrections in matched next-to-leading-order shower event generators},
  year          = {2013},
  pages         = {094009},
  volume        = {87},
  archiveprefix = {arXiv},
  doi           = {10.1103/PhysRevD.87.094009},
  eprint        = {1212.6164},
  primaryclass  = {hep-ph},
  reportnumber  = {DESY-12-166, OUTP-12-19P},
}

@Article{CMS:2014qtp,
  author        = {Khachatryan, Vardan and others},
  journal       = {Eur. Phys. J. C},
  title         = {Constraints on parton distribution functions and extraction of the strong coupling constant from the inclusive jet cross section in pp collisions at $\sqrt{s} = 7\,${TeV}},
  year          = {2015},
  number        = {6},
  pages         = {288},
  volume        = {75},
  archiveprefix = {arXiv},
  collaboration = {CMS},
  doi           = {10.1140/epjc/s10052-015-3499-1},
  eprint        = {1410.6765},
  primaryclass  = {hep-ex},
  reportnumber  = {CMS-SMP-12-028, CERN-PH-EP-2014-238},
}

@Article{Bierlich:2019rhm,
  author        = {Christian Bierlich and Andy Buckley and Jonathan Butterworth and Christian Holm Christensen and Louie Corpe and others},
  journal       = {SciPost Phys.},
  title         = {{R}obust {I}ndependent {V}alidation of {E}xperiment and {T}heory: {R}ivet version 3},
  year          = {2020},
  pages         = {026},
  volume        = {8},
  archiveprefix = {arXiv},
  doi           = {10.21468/SciPostPhys.8.2.026},
  eprint        = {1912.05451},
  primaryclass  = {hep-ph},
  reportnumber  = {MCnet-19-26},
}

@Article{Platzer:2022jny,
  author        = {Pl\"atzer, Simon},
  journal       = {JHEP},
  title         = {Colour evolution and infrared physics},
  year          = {2023},
  pages         = {126},
  volume        = {07},
  archiveprefix = {arXiv},
  doi           = {10.1007/JHEP07(2023)126},
  eprint        = {2204.06956},
  primaryclass  = {hep-ph},
  reportnumber  = {UWTHPH-2020-87, MCnet-22-07},
}

@Article{Czakon:2019tmo,
  author        = {Czakon, Michal and van Hameren, Andreas and Mitov, Alexander and Poncelet, Rene},
  journal       = {JHEP},
  title         = {Single-jet inclusive rates with exact color at $\mathcal{O}(\alpha_s^4)$},
  year          = {2019},
  pages         = {262},
  volume        = {10},
  archiveprefix = {arXiv},
  doi           = {10.1007/JHEP10(2019)262},
  eprint        = {1907.12911},
  primaryclass  = {hep-ph},
}

@Article{Bewick:2023tfi,
  author        = {Bewick, Gavin and others},
  journal       = {Eur. Phys. J. C},
  title         = {{Herwig 7.3 release note}},
  year          = {2024},
  number        = {10},
  pages         = {1053},
  volume        = {84},
  archiveprefix = {arXiv},
  doi           = {10.1140/epjc/s10052-024-13211-9},
  eprint        = {2312.05175},
  primaryclass  = {hep-ph},
  reportnumber  = {CERN-TH-2023-223, HERWIG-2023-01, KA-TP-28-2023, MCnet-23-19, IPPP/23/66},
}

@PhdThesis{Horzela2023,
  author    = {Horzela, Maximilian Maria},
  school    = {Karlsruher Institut für Technologie ({KIT})},
  title     = {Measurement of Triple-Differential {Z+Jet} Cross Sections with the {CMS} Detector at 13 {TeV} and Modelling of Large-Scale Distributed Computing Systems},
  year      = {2023},
  doi       = {10.5445/IR/1000165566},
  language  = {english},
  pagetotal = {281},
  publisher = {Karlsruher Institut für Technologie ({KIT})},
}

@Article{CDF:2010pdo,
  author        = {Aaltonen, T. and others},
  journal       = {Phys. Rev. D},
  title         = {Studying the underlying event in {D}rell-{Y}an and high transverse momentum jet production at the {T}evatron},
  year          = {2010},
  pages         = {034001},
  volume        = {82},
  archiveprefix = {arXiv},
  collaboration = {CDF},
  doi           = {10.1103/PhysRevD.82.034001},
  eprint        = {1003.3146},
  primaryclass  = {hep-ex},
  reportnumber  = {FERMILAB-PUB-10-053-E},
}

@Article{ParticleDataGroup:2022pth,
  author        = {Workman, R. L. and others},
  journal       = {PTEP},
  title         = {{R}eview of {P}article {P}hysics},
  year          = {2022},
  pages         = {083C01},
  volume        = {2022},
  collaboration = {Particle Data Group},
  comment       = {MZ = 91.1876 ± 0.0021; the 2024 value is: MZ = 91.1880 ± 0.0020},
  doi           = {10.1093/ptep/ptac097},
}

@Article{Gieseke:2012ft,
  author        = {Gieseke, Stefan and Rohr, Christian and Siodmok, Andrzej},
  journal       = {Eur. Phys. J. C},
  title         = {Colour reconnections in {H}erwig++},
  year          = {2012},
  pages         = {2225},
  volume        = {72},
  archiveprefix = {arXiv},
  doi           = {10.1140/epjc/s10052-012-2225-5},
  eprint        = {1206.0041},
  primaryclass  = {hep-ph},
  reportnumber  = {MCNET-12-06, KA-TP-17-2012, MAN-HEP-2012-03},
}

@Article{Ablat:2024uvg,
  author        = {Ablat, Alim and Dulat, Sayipjamal and Hou, Tie-Jiun and Huston, Joey and Nadolsky, Pavel and Sitiwaldi, Ibrahim and Xie, Keping and Yuan, C. -P.},
  journal       = {Phys. Rev. D},
  title         = {Impact of {LHC} precision measurements of inclusive jet and dijet production on the {CTEQ-TEA} global {PDF} fit},
  year          = {2025},
  number        = {3},
  pages         = {036033},
  volume        = {111},
  archiveprefix = {arXiv},
  collaboration = {CTEQ-TEA},
  doi           = {10.1103/PhysRevD.111.036033},
  eprint        = {2412.00350},
  primaryclass  = {hep-ph},
}

@Article{Hou:2019efy,
  author        = {Hou, Tie-Jiun and others},
  journal       = {Phys. Rev. D},
  title         = {New {CTEQ} global analysis of quantum chromodynamics with high-precision data from the {LHC}},
  year          = {2021},
  number        = {1},
  pages         = {014013},
  volume        = {103},
  archiveprefix = {arXiv},
  doi           = {10.1103/PhysRevD.103.014013},
  eprint        = {1912.10053},
  primaryclass  = {hep-ph},
  reportnumber  = {MSUHEP-19-025, PITT-PACC-1911, SMU-HEP-19-03},
}

@Article{CMS:2017jfq,
  author        = {Sirunyan, Albert M and others},
  journal       = {Eur. Phys. J. C},
  title         = {Measurement of the triple-differential dijet cross section in proton-proton collisions at $\sqrt{s} = 8\,${TeV} and constraints on parton distribution functions},
  year          = {2017},
  number        = {11},
  pages         = {746},
  volume        = {77},
  archiveprefix = {arXiv},
  collaboration = {CMS},
  doi           = {10.1140/epjc/s10052-017-5286-7},
  eprint        = {1705.02628},
  primaryclass  = {hep-ex},
  reportnumber  = {CMS-SMP-16-011, CERN-EP-2017-061},
}

@Article{Aad:2008zzm,
  author        = {Aad, G. and others},
  journal       = {JINST},
  title         = {The {ATLAS} {E}xperiment at the {CERN} {L}arge {H}adron {C}ollider},
  year          = {2008},
  pages         = {S08003},
  volume        = {3},
  collaboration = {ATLAS},
  comment       = {ATLAS detector reference},
  doi           = {10.1088/1748-0221/3/08/S08003},
}

@Article{Bellm:2015jjp,
  author        = {Bellm, Johannes and others},
  journal       = {Eur. Phys. J. C},
  title         = {Herwig 7.0 / {H}erwig++ 3.0 release note},
  year          = {2015},
  number        = {4},
  pages         = {196},
  volume        = {76},
  archiveprefix = {arXiv},
  doi           = {10.1140/epjc/s10052-016-4018-8},
  eprint        = {1512.01178},
  primaryclass  = {hep-ph},
  reportnumber  = {CERN-PH-TH-2015-289, MAN-HEP-2015-15, IFJPAN-IV-2015-13, HERWIG-2015-01, KA-TP-18-2015, DCPT-15-142, MCNET-15-28, IPPP-15-71},
}

@Article{Abramowicz:2015mha,
  author        = {Abramowicz, H. and others},
  journal       = {Eur. Phys. J. C},
  title         = {Combination of measurements of inclusive deep inelastic ${e^{\pm}p}$ scattering cross sections and {QCD} analysis of {HERA} data},
  year          = {2015},
  pages         = {580},
  volume        = {75},
  archiveprefix = {arXiv},
  collaboration = {H1 and ZEUS},
  comment       = {HERAPDF 2.0 reference},
  doi           = {10.1140/epjc/s10052-015-3710-4},
  eprint        = {1506.06042},
  primaryclass  = {hep-ex},
  reportnumber  = {DESY-15-039},
}

@Article{Gieseke:2018gff,
  author        = {Gieseke, Stefan and Kirchgaeßer, Patrick and Plätzer, Simon and Siodmok, Andrzej},
  journal       = {JHEP},
  title         = {Colour reconnection from soft gluon evolution},
  year          = {2018},
  pages         = {149},
  volume        = {11},
  archiveprefix = {arXiv},
  doi           = {10.1007/JHEP11(2018)149},
  eprint        = {1808.06770},
  primaryclass  = {hep-ph},
  reportnumber  = {KA-TP-23-2018, UWTHPH-2018-23, HERWIG-2018-02, MCnet-18-22},
}

@Article{Sherpa:2019gpd,
  author        = {Bothmann, Enrico and others},
  journal       = {SciPost Phys.},
  title         = {{E}vent generation with {S}herpa 2.2},
  year          = {2019},
  number        = {3},
  pages         = {034},
  volume        = {7},
  archiveprefix = {arXiv},
  collaboration = {Sherpa},
  doi           = {10.21468/SciPostPhys.7.3.034},
  eprint        = {1905.09127},
  primaryclass  = {hep-ph},
  reportnumber  = {FERMILAB-PUB-19-218-T, SLAC-PUB-17433, IPPP/19/42, MCNET-19-11},
}

@Article{Cacciari:2008gp,
  author        = {Cacciari, Matteo and Salam, Gavin P. and Soyez, Gregory},
  journal       = {JHEP},
  title         = {The {anti-$k_t$} jet clustering algorithm},
  year          = {2008},
  pages         = {063},
  volume        = {04},
  archiveprefix = {arXiv},
  comment       = {anti-kT Jet Algorithm Reference},
  doi           = {10.1088/1126-6708/2008/04/063},
  eprint        = {0802.1189},
  primaryclass  = {hep-ph},
}

@Article{Cacciari:2011ma,
  author        = {Cacciari, Matteo and Salam, Gavin P. and Soyez, Gregory},
  journal       = {Eur. Phys. J. C},
  title         = {{F}ast{J}et user manual},
  year          = {2012},
  pages         = {1896},
  volume        = {72},
  archiveprefix = {arXiv},
  comment       = {fastjet Reference},
  doi           = {10.1140/epjc/s10052-012-1896-2},
  eprint        = {1111.6097},
  primaryclass  = {hep-ph},
  reportnumber  = {CERN-PH-TH-2011-297},
}

@Article{NNPDF:2021njg,
  author        = {Ball, Richard D. and others},
  journal       = {Eur. Phys. J. C},
  title         = {The path to proton structure at 1\% accuracy},
  year          = {2022},
  month         = {9},
  number        = {5},
  pages         = {428},
  volume        = {82},
  archiveprefix = {arXiv},
  collaboration = {NNPDF},
  doi           = {10.1140/epjc/s10052-022-10328-7},
  eprint        = {2109.02653},
  primaryclass  = {hep-ph},
  reportnumber  = {Edinburgh 2021/12, Nikhef-2021-013, TIF-UNIMI-2021-11},
}

@Article{ATLAS:2019ocl,
  author        = {Aad, Georges and others},
  journal       = {Eur. Phys. J. C},
  title         = {Measurement of distributions sensitive to the underlying event in inclusive {Z} boson production in pp collisions at $\sqrt{s} = 13\,${TeV} with the {ATLAS} detector},
  year          = {2019},
  number        = {8},
  pages         = {666},
  volume        = {79},
  archiveprefix = {arXiv},
  collaboration = {ATLAS},
  doi           = {10.1140/epjc/s10052-019-7162-0},
  eprint        = {1905.09752},
  primaryclass  = {hep-ex},
  reportnumber  = {CERN-EP-2019-064},
}

@Article{Chen:2022tpk,
  author        = {Chen, X. and Gehrmann, T. and Glover, E. W. N. and Huss, A. and Mo, J.},
  journal       = {JHEP},
  title         = {{NNLO QCD} corrections in full colour for jet production observables at the {LHC}},
  year          = {2022},
  pages         = {025},
  volume        = {09},
  archiveprefix = {arXiv},
  comment       = {TODO: Jets at NNLO, full colour (FC) vs. leading colour (LC)},
  doi           = {10.1007/JHEP09(2022)025},
  eprint        = {2204.10173},
  primaryclass  = {hep-ph},
  reportnumber  = {ZU-TH 11/22, KA-TP-07-2022, IPPP/22/20, P3H-22-037, CERN-TH-2022-067},
}

@Article{CMS:2012oqb,
  author        = {Chatrchyan, Serguei and others},
  journal       = {Eur. Phys. J. C},
  title         = {Measurement of the underlying event in the {D}rell-{Y}an process in proton-proton collisions at $\sqrt{s}=7\,${TeV}},
  year          = {2012},
  pages         = {2080},
  volume        = {72},
  archiveprefix = {arXiv},
  collaboration = {CMS},
  doi           = {10.1140/epjc/s10052-012-2080-4},
  eprint        = {1204.1411},
  primaryclass  = {hep-ex},
  reportnumber  = {CMS-QCD-11-012, CERN-PH-EP-2012-085},
}

@Article{Platzer:2011bc,
  author        = {Pl\"atzer, Simon and Gieseke, Stefan},
  journal       = {Eur. Phys. J. C},
  title         = {Dipole showers and automated {NLO} matching in {H}erwig++},
  year          = {2012},
  pages         = {2187},
  volume        = {72},
  archiveprefix = {arXiv},
  doi           = {10.1140/epjc/s10052-012-2187-7},
  eprint        = {1109.6256},
  primaryclass  = {hep-ph},
  reportnumber  = {DESY-11-162, KA-TP-24-2011, HERWIG-11-01, MCNET-11-24},
}

@Misc{marcel_rieger_2024_13952360,
  author    = {Rieger, Marcel and Rath, Yannik and Noll, Dennis and Geiger, Lukas and Fackeldey, Peter and others},
  month     = oct,
  note      = {v0.1.19},
  title     = {riga/law},
  year      = {2024},
  doi       = {10.5281/zenodo.13952360},
  publisher = {Zenodo},
}

@Misc{horzela_2024_14530596,
  author    = {Horzela, Maximilian and Verstege, Cedric},
  month     = dec,
  note      = {v1.1.0},
  title     = {{MCRun}: Scalable {MC} Event Generation and {R}ivet Analysis},
  year      = {2024},
  doi       = {10.5281/zenodo.14530596},
  publisher = {Zenodo},
}

@Article{Gleisberg:2008ta,
  author        = {Gleisberg, T. and Höche, Stefan. and Krauss, F. and Schönherr, M. and Schumann, S. and others},
  journal       = {JHEP},
  title         = {Event generation with {SHERPA} 1.1},
  year          = {2009},
  pages         = {007},
  volume        = {02},
  archiveprefix = {arXiv},
  comment       = {SHERPA reference},
  doi           = {10.1088/1126-6708/2009/02/007},
  eprint        = {0811.4622},
  primaryclass  = {hep-ph},
  reportnumber  = {FERMILAB-PUB-08-477-T, SLAC-PUB-13420, ZU-TH-17-08, DCPT-08-138, IPPP-08-69, EDINBURGH-2008-30, MCNET-08-14},
}

@Article{ATLAS:2014yqy,
  author        = {Aad, Georges and others},
  journal       = {Eur. Phys. J. C},
  title         = {Measurement of distributions sensitive to the underlying event in inclusive {Z}-boson production in $pp$ collisions at $\sqrt{s}=7\,${TeV} with the {ATLAS} detector},
  year          = {2014},
  number        = {12},
  pages         = {3195},
  volume        = {74},
  archiveprefix = {arXiv},
  collaboration = {ATLAS},
  doi           = {10.1140/epjc/s10052-014-3195-6},
  eprint        = {1409.3433},
  primaryclass  = {hep-ex},
  reportnumber  = {CERN-PH-EP-2014-162},
}

@Article{CDF:2001onq,
  author        = {Affolder, T. and others},
  journal       = {Phys. Rev. D},
  title         = {Charged jet evolution and the underlying event in proton-antiproton collisions at 1.8\,{TeV}},
  year          = {2002},
  pages         = {092002},
  volume        = {65},
  collaboration = {CDF},
  doi           = {10.1103/PhysRevD.65.092002},
  reportnumber  = {FERMILAB-PUB-01-211-E},
}

@Article{Bahr:2008pv,
  author        = {Bähr, M. and Gieseke, S. and Gigg, M. A. and Grellscheid, D. and Hamilton, K. and others},
  journal       = {Eur. Phys. J. C},
  title         = {Herwig++ physics and manual},
  year          = {2008},
  pages         = {639},
  volume        = {58},
  archiveprefix = {arXiv},
  comment       = {Herwig++ Reference},
  doi           = {10.1140/epjc/s10052-008-0798-9},
  eprint        = {0803.0883},
  primaryclass  = {hep-ph},
  reportnumber  = {CERN-PH-TH-2008-038, CAVENDISH-HEP-08-03, KA-TP-05-2008, DCPT-08-22, IPPP-08-11, CP3-08-05},
}

@Article{Boughezal:2015ded,
  author        = {Boughezal, Radja and Campbell, John M. and Ellis, R. Keith and Focke, Christfried and Giele, Walter T. and Liu, Xiaohui and Petriello, Frank},
  journal       = {Phys. Rev. Lett.},
  title         = {{Z}-{B}oson {P}roduction in {A}ssociation with a {J}et at {N}ext-{T}o-{N}ext-{T}o-{L}eading {O}rder in {P}erturbative {QCD}},
  year          = {2016},
  pages         = {152001},
  volume        = {116},
  archiveprefix = {arXiv},
  comment       = {Z+1jet NNLO Njettiness},
  doi           = {10.1103/PhysRevLett.116.152001},
  eprint        = {1512.01291},
  primaryclass  = {hep-ph},
}

@Article{Ridder:2016nkl,
  author        = {Gehrmann-De Ridder, Aude and Gehrmann, T. and Glover, E. W. N. and Huss, A. and Morgan, T. A.},
  journal       = {JHEP},
  title         = {The {NNLO} {QCD} corrections to {Z} boson production at large transverse momentum},
  year          = {2016},
  pages         = {133},
  volume        = {07},
  archiveprefix = {arXiv},
  comment       = {TODO: NNLOJET Z+jet},
  doi           = {10.1007/JHEP07(2016)133},
  eprint        = {1605.04295},
  primaryclass  = {hep-ph},
  reportnumber  = {IPPP-16-39, NSF-KITP-16-067, ZU-TH-18-16},
}

@Article{Chatrchyan:2008aa,
  author        = {Chatrchyan, S. and others},
  journal       = {JINST},
  title         = {The {CMS} experiment at the {CERN} {LHC}},
  year          = {2008},
  pages         = {S08004},
  volume        = {3},
  collaboration = {CMS},
  comment       = {CMS detector reference},
  doi           = {10.1088/1748-0221/3/08/S08004},
}

@Article{Hoang:2024nqi,
  author        = {Hoang, Andr{\'e} H. and Jin, Oliver L. and Pl{\"a}tzer, Simon and Samitz, Daniel},
  journal       = {JHEP},
  title         = {{Matching hadronization and perturbative evolution: the cluster model in light of infrared shower cutoff dependence}},
  year          = {2025},
  pages         = {005},
  volume        = {07},
  archiveprefix = {arXiv},
  doi           = {10.1007/JHEP07(2025)005},
  eprint        = {2404.09856},
  primaryclass  = {hep-ph},
  reportnumber  = {UWThPh-2023-23, MCnet-24-05},
}

@Article{Christiansen:2015yca,
  author        = {Christiansen, Jesper R. and Sj\"ostrand, Torbj\"orn},
  journal       = {Eur. Phys. J. C},
  title         = {Color reconnection at future e$^{+}$\,e$^{-}$ colliders},
  year          = {2015},
  number        = {9},
  pages         = {441},
  volume        = {75},
  archiveprefix = {arXiv},
  doi           = {10.1140/epjc/s10052-015-3674-4},
  eprint        = {1506.09085},
  primaryclass  = {hep-ph},
  reportnumber  = {LU-TP-15-25, MCNET-15-15},
}

@Article{CMS:2017ngy,
  author        = {Sirunyan, A. M. and others},
  journal       = {JHEP},
  title         = {Measurement of the underlying event activity in inclusive {Z} boson production in proton-proton collisions at $\sqrt{s}=13\,${TeV}},
  year          = {2018},
  pages         = {032},
  volume        = {07},
  archiveprefix = {arXiv},
  collaboration = {CMS},
  doi           = {10.1007/JHEP07(2018)032},
  eprint        = {1711.04299},
  primaryclass  = {hep-ex},
  reportnumber  = {CMS-FSQ-16-008, CERN-EP-2017-249},
}

@Article{Christiansen:2015yqa,
  author        = {Christiansen, Jesper R. and Skands, Peter Z.},
  journal       = {JHEP},
  title         = {String formation beyond leading colour},
  year          = {2015},
  pages         = {003},
  volume        = {08},
  archiveprefix = {arXiv},
  doi           = {10.1007/JHEP08(2015)003},
  eprint        = {1505.01681},
  primaryclass  = {hep-ph},
  reportnumber  = {COEPP-MN-15-1, LU-TP-15-16, MCNET-15-09, COEPP-MN-15-1, LU-TP-15-16, MCNET-15-09},
}

@Article{CMS:2023fix,
  author        = {Hayrapetyan, Aram and others},
  journal       = {Eur. Phys. J. C},
  title         = {Measurement of multidifferential cross sections for dijet production in proton-proton collisions at $\sqrt{s} = 13\,${TeV}},
  year          = {2025},
  month         = {12},
  number        = {1},
  pages         = {72},
  volume        = {85},
  archiveprefix = {arXiv},
  collaboration = {CMS},
  doi           = {10.1140/epjc/s10052-024-13606-8},
  eprint        = {2312.16669},
  primaryclass  = {hep-ex},
  reportnumber  = {CMS-SMP-21-008, CERN-EP-2023-257},
}

@Article{Currie:2017eqf,
  author        = {Currie, James and Gehrmann-De Ridder, Aude and Gehrmann, Thomas and Glover, E. W. N. and Huss, Alexander and Pires, Joao},
  journal       = {Phys. Rev. Lett.},
  title         = {{P}recise {P}redictions for {D}ijet {P}roduction at the {LHC}},
  year          = {2017},
  number        = {15},
  pages         = {152001},
  volume        = {119},
  archiveprefix = {arXiv},
  comment       = {NNLOJET dijet reference},
  doi           = {10.1103/PhysRevLett.119.152001},
  eprint        = {1705.10271},
  primaryclass  = {hep-ph},
  reportnumber  = {IPPP-17-45, ZU-TH-13-17, MPP-2017-107},
}

@Article{Cridge:2023ozx,
  author        = {Cridge, T. and Harland-Lang, L. A. and Thorne, R. S.},
  journal       = {Eur. Phys. J. C},
  title         = {The impact of {LHC} jet and {$Z\,p_T$} data at up to approximate {N${}^3$LO} order in the {MSHT} global {PDF} fit},
  year          = {2024},
  number        = {4},
  pages         = {446},
  volume        = {84},
  archiveprefix = {arXiv},
  doi           = {10.1140/epjc/s10052-024-12771-0},
  eprint        = {2312.12505},
  primaryclass  = {hep-ph},
  reportnumber  = {DESY-23-185},
}

@Article{CDF:2004jod,
  author        = {Acosta, D. and others},
  journal       = {Phys. Rev. D},
  title         = {Underlying event in hard interactions at the {F}ermilab {T}evatron $\bar{p}p$ collider},
  year          = {2004},
  pages         = {072002},
  volume        = {70},
  archiveprefix = {arXiv},
  collaboration = {CDF},
  doi           = {10.1103/PhysRevD.70.072002},
  eprint        = {hep-ex/0404004},
  reportnumber  = {FERMILAB-PUB-04-041-E},
}

@Article{D0:2011jpq,
  author        = {Abazov, Victor Mukhamedovich and others},
  journal       = {Phys. Rev. D},
  title         = {Measurement of the inclusive jet cross section in $p \bar {p}$ collisions at $\sqrt{s} = 1.96$ {TeV}},
  year          = {2012},
  pages         = {052006},
  volume        = {85},
  archiveprefix = {arXiv},
  collaboration = {D0},
  doi           = {10.1103/PhysRevD.85.052006},
  eprint        = {1110.3771},
  primaryclass  = {hep-ex},
  reportnumber  = {FERMILAB-PUB-11-547-E},
}

@Article{ATLAS:2025ifq,
  author        = {Aad, Georges and others},
  title         = {Measurement of inclusive dijet cross-sections in proton-proton collisions at $\sqrt{s} = 13$ {TeV} with the {ATLAS} detector},
  year          = {2025},
  month         = {12},
  archiveprefix = {arXiv},
  collaboration = {ATLAS},
  eprint        = {2512.19073},
  primaryclass  = {hep-ex},
  reportnumber  = {CERN-EP-2025-283},
}

@Article{CDF:2007ykk,
  author        = {Aaltonen, T. and others},
  journal       = {Phys. Rev. Lett.},
  title         = {Measurement of Inclusive Jet Cross Sections in $Z/\gamma^*(\to e^{+} e^{-})$ + jets Production in $p \bar{p}$ Collisions at $\sqrt{s} = 1.96$ {TeV}},
  year          = {2008},
  pages         = {102001},
  volume        = {100},
  archiveprefix = {arXiv},
  collaboration = {CDF},
  doi           = {10.1103/PhysRevLett.100.102001},
  eprint        = {0711.3717},
  primaryclass  = {hep-ex},
  reportnumber  = {FERMILAB-PUB-07-627-E},
}


\end{document}